# Reliability-Aware Hybrid-K Ensemble Selection for Cervical Cytology Classification: Integrating Discrimination, Calibration, and Selective Prediction

Nisreen Albzour, Sarah S. Lam

School of Systems Science and Industrial Engineering, Binghamton University, Binghamton, NY 13902, USA

Email: nalbzour@binghamton.edu, sarahlam@binghamton.edu

*Corresponding author: nalbzour@binghamton.edu*

## Abstract

High classification accuracy alone is insufficient for deploying deep learning models in clinical image analysis, where calibrated confidence and reliable uncertainty handling are essential. This study proposes a reliability-aware Hybrid-K ensemble selection framework for multiclass cervical cytology classification using the SIPaKMeD dataset. Nine candidate deep learning architectures were trained under a shared protocol using a fixed stratified five-fold partition, with training repeated under three independent training seeds. After post-hoc temperature scaling, models were evaluated using eight metrics spanning classification performance, calibration, probabilistic reliability, and selective prediction: macro-F1, accuracy, area under the receiver operating characteristic curve (AUROC), expected calibration error (ECE), worst-class ECE (WC-ECE), area under the risk-coverage curve (AURC), Brier score, and negative log-likelihood (NLL). Candidate models were ranked using an equal-weight reliability-aware composite score, and Hybrid-K ensembles were constructed from the top-ranked models using soft voting. Robustness was assessed using 5,000 Dirichlet-sampled metric-weight vectors, leave-one-metric-out sensitivity analysis, and corrected repeated cross-validation paired statistical testing across 15 fold-by-seed evaluations (5 fixed folds × 3 training seeds). The final Hybrid-2 ensemble, composed of Swin-Tiny and TinyViT-5M, reduced AURC by 43%, NLL by 17%, and worst-class ECE by 36% compared with the best individual model. It was selected in 96.8% of random weighting scenarios, remained unchanged across all leave-one-metric-out ablations, and improved the full composite score. Those results demonstrate the value of the proposed Hybrid-K framework, although the per-metric gains over the best individual model were not statistically significant under the corrected repeated cross-validation paired t-test with Holm-Bonferroni correction (all Holm-adjusted $p \geq 0.168$). Because the post-hoc calibration procedure did not use a fully independent calibration set, the calibration-dependent quantities reported here should be interpreted as internal exploratory estimates. These results show that reliability-aware ensemble selection can identify a compact ensemble that is robust to alternative metric weightings and improves reliability point estimates for cervical cytology classification under internal validation on a single dataset.

## 1. Introduction

Cervical cancer is one of the most common malignancies affecting women worldwide and a leading cause of cancer-related mortality in low- and middle-income countries, despite being largely preventable through early detection [1]. Cytology-based screening programs, in which exfoliated cervical cells are examined microscopically, enable the identification of precancerous lesions before progression to invasive disease. Early computer-assisted approaches to digital image processing of Pap smear images laid the groundwork for these automated screening pipelines [2]. Manual cytological examination is, however, labor-intensive, subject to inter-observer variability, and difficult to scale in settings where trained cytopathologists are scarce. Automated image classification systems driven by deep learning offer a pathway toward more consistent, high-throughput cervical cell analysis and have the potential to serve as an effective second-reader tool in clinical screening workflows. Comparable automation efforts have been pursued in other cytology- and hematology-based cancer detection tasks, such as leukemia classification, where hyperautomation pipelines have been proposed to reduce the burden of manual review [3].

Deep learning has demonstrated strong performance on cervical cytology image classification benchmarks, with convolutional neural networks (CNNs), vision transformers, and hybrid architectures achieving high accuracy on publicly available datasets such as SIPaKMeD [4]. That landscape was synthesized in 2024 in a systematic review of segmentation and classification techniques for Pap smear images [5]. Existing studies typically evaluate these models using discrimination metrics (accuracy and macro-averaged F1 in particular) and optimize model selection to maximize predictive performance on held-out test sets. However, high classification accuracy alone is insufficient for clinical deployment. In safety-critical applications, a model's predicted class probabilities must be reliably calibrated: a model that assigns 90% confidence to a prediction should be correct approximately 90% of the time [6], a property that is inconsistently achieved across medical image classification tasks more broadly [7]. Models that are overconfident or systematically miscalibrated can mislead downstream clinical decision-making, particularly in scenarios where uncertain predictions should be flagged for expert review rather than acted upon automatically [8]. That concern is not specific to image classification: comparable reliability requirements arise in broader clinical prediction tasks, such as post-stroke functional outcome prediction, where model performance must remain robust under real-world data and feature constraints [9]. Concerns about whether deep learning diagnoses can be trusted under real-world conditions have been raised broadly across medical imaging, including domain-shift-induced failures in chest radiograph classification [10].

Alongside calibration, selective prediction is an important reliability property in clinical image classification. A model that can identify and abstain from its most uncertain predictions, referring those cases to a specialist while classifying the remainder with high confidence, provides greater practical utility than a model that assigns confident predictions indiscriminately. Metrics such as ECE, worst-class ECE (WC-ECE), AURC [11], Brier score, and NLL capture these complementary dimensions of reliability and probabilistic quality. Despite their importance, those metrics rarely appear in the literature as primary selection criteria in deep learning studies on cervical cytology or in systematic frameworks for jointly optimizing classification performance and reliability when selecting models or ensembles for this task. Lambert et al. [8], in a systematic review of uncertainty quantification across 218 deep learning studies in medical image analysis, explicitly identified the mismatch between high model accuracy and trustworthy predictive confidence as a central open challenge in clinical AI deployment, directly motivating the reliability-aware evaluation framework developed in this work. This work builds on earlier, narrower-scope reviews of uncertainty estimation in medical image classification, such as a systematic review covering 22 papers [12], reflecting the field's rapidly growing but fragmented attention to predictive reliability.

Ensemble methods, which combine predictions from multiple independently trained models, offer a principled strategy for improving classification performance and probability estimate reliability. Ensemble averaging reduces variance in predicted probabilities and can improve calibration relative to individual constituent models. The central challenge in ensemble construction is determining how many models to include and which models to select: adding more members increases computational complexity without guaranteeing proportional gains in reliability. Most existing ensemble selection strategies for medical image classification optimize for a single criterion (typically accuracy or F1) and do not incorporate reliability-oriented metrics into the selection process. A framework that jointly accounts for classification performance, calibration error, selective prediction behavior, and probabilistic scoring when constructing and validating an ensemble is needed. Establishing the framework on a well-characterized benchmark is a necessary first step. Reliability-aware validation also motivates assessment beyond a single dataset. Recent work on breast cancer survival prediction demonstrates that calibration and generalization are best assessed on data drawn from a different distribution than the one used for model development [13]. Reliability-oriented evaluation beyond accuracy has similarly been examined for imbalance-handling strategies outside medical imaging [14]. Complementary evidence from transformer-based text classification shows that post-hoc temperature scaling combined with uncertainty-aware selective prediction improves calibration and risk-coverage behavior under class imbalance without degrading accuracy [15].

This paper proposes a five-stage reliability-aware ensemble selection framework for cervical cytology image classification on the SIPaKMeD dataset. Nine candidate architectures spanning CNNs, vision transformers, and hybrid CNN–transformer designs are trained under a shared experimental recipe, calibrated with post-hoc temperature scaling, and evaluated using an eight-metric composite score that integrates three classification metrics (macro-F1, accuracy, AUROC) with five reliability metrics (ECE, WC-ECE, AURC, Brier score, NLL). Candidate models are ranked by this composite score, and Hybrid-K ensembles are formed from the top-ranked individual models using soft voting. The selected ensemble is then validated through three complementary analyses: (i) a Dirichlet-based robustness analysis using 5,000 randomly sampled metric-weight vectors to assess whether the ensemble selection is robust under alternative weighting assumptions, (ii) a leave-one-metric-out (LOMO) sensitivity analysis to determine which metrics drive ensemble membership, and (iii) corrected repeated cross-validation paired t-testing (Nadeau and Bengio) with Holm-Bonferroni correction across 15 fold-by-seed evaluations (5 fixed folds × 3 training seeds) to assess whether observed gains are statistically distinguishable from fold-level variability. The primary contributions of this work are threefold. First, a reliability-aware composite score is introduced that unifies discrimination, calibration, and selective-prediction metrics into a single model-selection criterion, moving selection beyond accuracy-only optimization. Second, a Hybrid-K ensemble construction and selection procedure is developed based on this composite score. Third, a validation methodology is established that combines Dirichlet-based robustness analysis, leave-one-metric-out sensitivity analysis, and corrected repeated cross-validation paired testing to determine whether the selected ensemble is robust to alternative metric weightings and statistically defensible. Frameworks that jointly optimize classification performance and reliability for ensemble selection in cervical cytology classification remain scarce, and this work contributes one such framework.

The remainder of this paper is organized as follows: Section 2 reviews related work on deep learning for cervical cytology classification, ensemble methods, calibration and selective prediction, and multi-criteria model selection. Section 3 describes the experimental methodology in detail, including dataset preparation and cross-validation design, candidate model training, calibration and reliability evaluation, composite ranking, ensemble construction, and the three validation analyses. Section 4 presents and discusses the results across all five evaluation stages. Section 5 concludes the paper and outlines directions for future work.

## 2. Related Work

This section reviews prior work along the dimensions most relevant to reliability-aware ensemble selection for cervical cytology. Deep learning methods for cervical cell classification (Section 2.1) and ensemble strategies in medical image classification (Section 2.2) are surveyed first, followed by the reliability properties that motivate the present framework—calibration and selective prediction (Section 2.3) and generalization under domain shift together with multi-criteria model selection (Section 2.4). Finally, Section 2.5 positions the present study relative to the closest prior work and highlights the gap it addresses. Figure 1 summarizes the organization of this review across the five dimensions.

More broadly, this study is situated within the long-standing Artificial Intelligence in Medicine literature on machine learning for medical diagnosis and clinical decision support, spanning early foundational work on knowledge-based medical reasoning and bounded rationality, historical reviews of machine learning for medical diagnosis, and broader retrospectives on the field's maturation and research themes [16–20]. These works establish the context for AI-based medical decision support generally, rather than addressing cervical cytology classification or reliability-aware ensemble selection specifically.

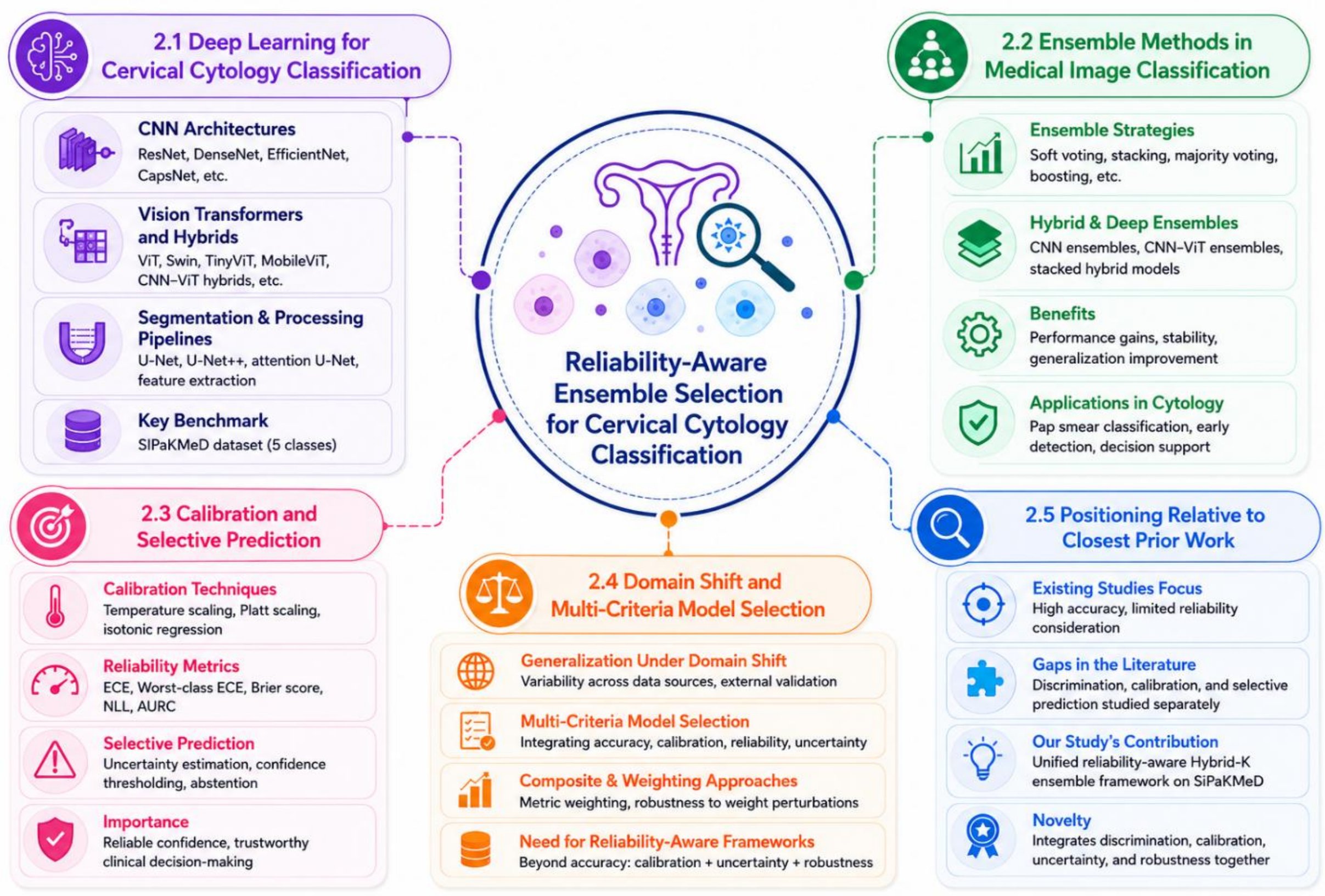

*Figure 1. Overview of the related work reviewed in this study, organized into five themes: deep learning for cervical cytology classification (Section 2.1), ensemble methods in medical image classification (Section 2.2), calibration and selective prediction (Section 2.3), generalization under domain shift and multi-criteria model selection (Section 2.4), and positioning relative to the closest prior work (Section 2.5).*

## 2.1 Deep Learning for Cervical Cytology Classification

Earlier work on automated Pap smear analysis explored classical feature-based classification and segmentation pipelines using shape, texture, and color features [21, 22], later reviewed comprehensively alongside the field's transition toward machine learning-based screening [23]; more recent work has continued to refine dedicated fusion-based CNN architectures specifically for Pap smear classification [24]. Deep learning has become the dominant approach for automated cervical cell classification, with the SIPaKMeD dataset [4] serving as the principal five-class benchmark since its introduction. Transfer learning from ImageNet-pretrained CNNs (including EfficientNet, ResNet, and DenseNet variants) has consistently yielded high accuracy on SIPaKMeD, with top-performing models exceeding 97% five-class accuracy; systematic comparisons of such transfer-learning backbones and dedicated CNN architectures for Pap smear classification reinforce this pattern [25, 26]. More recently, vision transformer (ViT) architectures and CNN–ViT hybrids have entered the field. Pacal and Kılıçarslan [27], in the most comprehensive evaluation to date, benchmarked 40 CNN and more than 20 ViT architectures on SIPaKMeD, finding that both families achieve competitive five-class performance when fine-tuned with strong augmentation. Consistent with this, a systematic evaluation of vision transformers for cervical cancer classification (combining hyperparameter optimization, statistical validation, and clinical interpretability) has shown that transformer backbones can be benchmarked against CNNs under rigorous, reproducible protocols rather than accuracy alone [28], and a focused reproducible benchmark of ViT-Tiny against CNN baselines on cervical cell classification found that such lightweight transformers can match established CNN backbones while offering deployment-efficiency advantages [29]. Kılıç and Kılıç [30] extended this line of work with multi-scale cytomorphological feature integration for high-accuracy classification, and fused CNN–ViT and hybrid transformer–CNN ensemble architectures have likewise been reported to gain accuracy over either family alone [31, 32]. Related work has combined U-Net-based segmentation with downstream classification on the Herlev Pap smear dataset for cervical cancer detection [33]. Despite these accuracy gains, the more recent deep learning models carry practical drawbacks. Transformer and hybrid architectures are typically more data- and compute-intensive than earlier feature-based pipelines, depend on large volumes of labeled cytology images that are costly to annotate, and remain harder to interpret for clinical end users. High-capacity models are also prone to overconfident, poorly calibrated predictions and to performance degradation under scanner- and population-induced domain shift, so strong accuracy on a single benchmark does not by itself guarantee reliable clinical behavior. A consistent limitation of this body of work is that model selection relies almost exclusively on accuracy and F1, with calibration, probabilistic scoring, and selective prediction properties largely unexamined.

### 2.2 Ensemble Methods in Medical Image Classification

Ensemble methods in medical image classification have been applied primarily as accuracy-boosting strategies. Soft voting (averaging class probability vectors across constituent models before taking the argmax) is the most common aggregation rule and is consistently competitive with learned combination strategies at lower complexity. In cervical cytology specifically, a range of ensemble strategies have been proposed that differ mainly in how member predictions are combined: from soft voting over CNN and transformer backbones, through fuzzy-rank and majority-vote aggregation, to stacking with explainable-AI attribution [34–38]. A smaller number of studies move beyond pure accuracy: Sreelatha and Shivashetty [39] incorporated uncertainty-aware prediction ranking directly into a deep ensemble for Pap smear classification, one of the few works in this domain to combine ensembling with an explicit uncertainty signal. These studies demonstrate consistent accuracy gains over individual models, but none systematically examine how ensemble size and composition trade off against calibration or selective-prediction quality beyond accuracy. The question of how to select the minimum ensemble that achieves reliable calibration and selective prediction, without simply adding models until performance saturates, has not been studied systematically in this domain.

Within Artificial Intelligence in Medicine specifically, prior work has examined ensemble-based medical decision support from complementary perspectives, including case-by-case explanation of ensemble outputs for clinical users [40] and ensemble-based diagnostic classification across disease domains [41]. These studies support the broader relevance of ensemble-based decision support in medicine, but they do not address reliability-aware Hybrid-K ensemble selection for cervical cytology using calibration, selective prediction, and robustness analysis jointly.

### 2.3 Calibration and Selective Prediction

Confidence calibration (the alignment between a model's predicted class probabilities and its empirical correctness frequencies) is a critical requirement for clinical AI deployment. Guo et al. [6] demonstrated that modern deep neural networks are systematically overconfident and introduced temperature scaling, a single-parameter post-hoc method that re-scales logit magnitudes without altering predicted class labels, as a simple and effective remedy. Lambert et al. [8], in a systematic review of uncertainty quantification across 218 deep learning studies in medical image analysis, confirmed that overconfidence is pervasive across modalities and tasks, and identified the absence of principled uncertainty evaluation as a key barrier to trustworthy clinical deployment. Subsequent work has extended calibration methodology specifically within medical imaging, spanning empirical comparisons of post-hoc methods, weight-scaling alternatives to temperature scaling, noise-robust calibration under label noise, and trainable schemes that learn calibrated confidence jointly with the classification objective rather than as a post-hoc step [42–46]. Calibration metrics including ECE, WC-ECE, Brier score, and NLL capture distinct failure modes of poorly calibrated models. Despite their established importance, these metrics are seldom used as primary model selection criteria in cervical cytology classification research.

Selective prediction provides a complementary reliability lens: rather than forcing a classification on every input, a well-designed model can abstain on uncertain cases and refer them to a human reviewer. Geifman and El-Yaniv [11] formalized the risk–coverage framework for deep networks, showing that softmax-based confidence thresholding yields smooth risk–coverage tradeoffs, and introduced (AURC) as a single-number summary of selective prediction quality. Zhou et al. [47] recently provided a rigorous population-level characterization of AURC and its finite-sample estimators, clarifying statistical properties that had previously been used only heuristically. Calibration-aware and selective-recalibration procedures have extended selective classification so that the retained (non-abstained) predictions remain well calibrated rather than merely more accurate [48, 49]. In a clinical setting, Raghu et al. [50] showed that models can be trained to predict when a human second opinion is warranted, directly operationalizing the referral use case that motivates selective prediction in medicine. Hong et al. [51] combined ensembling with dynamic gating specifically for trust-aware rejection in medical image diagnostics, integrating selective prediction directly into an ensemble architecture rather than applying it as a downstream post-processing step. Lakshminarayanan et al. [52] showed that averaging the softmax outputs of multiple independently trained networks (a 'deep ensemble') not only improves discriminative accuracy but also substantially improves predictive uncertainty estimates, yielding lower AURC and better calibration than any individual constituent model. In cervical cytology specifically, a companion study [53] evaluated selective prediction and uncertainty-aware referral for Pap smear classification on the Herlev dataset, where a soft-voting ensemble of Swin-Tiny and TinyViT-5M halved AURC relative to the best single configuration and substantially extended the coverage at which zero errors were made, even though the accuracy and macro-F1 differences were not statistically significant. These results establish that ensemble construction, calibration, and selective prediction are complementary mechanisms for improving reliability, and motivate their joint evaluation within a unified multi-metric framework.

### 2.4 Generalization Under Domain Shift and Multi-Criteria Model Selection

A further reliability requirement, often neglected alongside calibration and selective prediction, is generalization under domain shift: a model's discrimination and calibration quality on its training distribution provide no guarantee of performance on data from a different acquisition protocol, scanner, or patient population. Coskun et al. [54] directly demonstrated this vulnerability for liquid-based cervical cytology, showing that scanner-induced domain shifts across multiple data sources substantially degrade cell-level classification performance even for models trained with deep transfer learning, and argued for explicit multi-source external validation rather than single-dataset benchmarking. More broadly across medical machine learning, external validation has been identified as a necessary but inconsistently applied practice, as models routinely perform better on data from their own training cohort than on new data because of overfitting or covariate shift [55]. This concern is compounded by taxonomy differences across the field's available public datasets: SIPaKMeD [4] uses a five-class cytomorphological scheme, Herlev [56] uses a seven-class dysplasia-severity scheme, and Mendeley LBC [57] uses the four-class Bethesda System, none of which map onto one another without an explicit crosswalk. Dey [58] recently proposed an ensemble learning approach specifically for categorizing the Bethesda System of reporting in cervical cytology, underscoring that Bethesda-based classification remains an active but separately studied taxonomy from the cytomorphological schemes used in datasets such as SIPaKMeD. Despite these known sources of distribution shift, external validation across independent cervical cytology datasets remains rare in the literature, and reliability metrics are essentially never evaluated under such shift alongside discrimination metrics.

Selecting among candidate models or ensembles when several reliability metrics disagree is itself a nontrivial decision problem, particularly when there is no principled way to fix a single metric weighting in advance. Stochastic multicriteria acceptability analysis (SMAA) [59] provides a principled framework for evaluating decision alternatives under this uncertainty: rank acceptability is computed by integrating over a distribution of weight vectors on the simplex, quantifying how frequently each alternative ranks first across all plausible weighting scenarios. Complementary methodological results establish how SMAA should be applied in practice: the symmetric Dirichlet distribution (concentration parameter $\alpha = 1$) provides a uniform prior over the weight simplex and is therefore the natural choice for weight-space exploration; a Monte Carlo sample of $N = 1/(3p^2)$ draws suffices to achieve a relative standard error of p in the estimated acceptability indices; and the framework admits a Bayesian interpretation that grounds its use for robustness analysis via rank acceptability [60–63]. Collectively, those results provide the methodological foundation for weight-space robustness analysis of how multi-criteria models and ensembles are selected. A related weight-sensitivity approach has recently been applied outside classification entirely, to multimodal CT-MRI registration, where a composite reliability score combining registration-quality signals was used to classify individual cases into risk categories and weight-sensitivity analysis identified which component metric dominated the composite score [64]; this illustrates that composite reliability scoring and weight-sensitivity analysis generalize across medical imaging tasks, though that work addresses registration quality control rather than classification ensemble selection.

### 2.5 Positioning Relative to Closest Prior Work

The closest prior work to the present study is Demirtaş Alpsalaz et al. [65], who proposed an explainability-aligned, reliability-weighted fuzzy ensemble for cervical cancer classification. Their study combined several deep CNN classifiers through a fuzzy-rank fusion scheme in which member contributions were weighted by explainability-derived reliability signals. It evaluated the resulting ensemble using conventional discrimination metrics (accuracy, F1, and AUROC) and calibration metrics (ECE, Brier score, and NLL), with post-hoc temperature scaling applied to improve calibration. Unlike the framework introduced here, that approach does not evaluate selective prediction or risk-coverage behavior (AURC) and does not assess robustness to alternative metric weightings through a composite, multi-metric selection criterion. In addition, the present study builds directly on a companion investigation [66], which applied a reliability-aware, calibration-focused ensemble analysis to liquid-based cervical cytology (the Mendeley LBC dataset) under the native four-class Bethesda taxonomy. Using the same lightweight architectures adopted here (Swin-Tiny, TinyViT-5M, and DenseNet121) together with soft-voting Hybrid-2 and Hybrid-3 ensembles, that study found that post-hoc temperature scaling markedly improved calibration (ECE, Brier score, and NLL) for every configuration while leaving discrimination essentially unchanged, and that ensemble size provided no consistent reliability benefit over the best individual model once all configurations were calibrated. That work, however, focused on class-imbalance mitigation within a single liquid-based dataset and did not introduce the composite, weighting-robust selection criterion spanning discrimination, calibration, and selective prediction that is central to the present framework. A second companion study [53] addressed selective prediction directly, applying the same lightweight backbones to the Herlev Pap smear dataset under a binary Normal-versus-Abnormal formulation with AURC as the primary endpoint. That study likewise found that soft-voting ensembling improved risk–coverage behavior without a statistically significant discrimination gain, but it examined a single binary task and did not construct or validate a composite, weighting-robust selection criterion. Taken together, this body of work reveals a consistent gap. The great majority of cervical cytology classification studies, whether based on individual CNNs, vision transformers, or ensembles, report accuracy and macro-F1 as their primary, and often only, evaluation criteria. Calibration and probabilistic reliability (ECE, WC-ECE, Brier score, NLL), selective prediction and risk–coverage behavior (AURC), external validation under dataset and domain shift, and principled, weighting-robust ensemble composition are each studied in isolation. These aspects are typically addressed in general medical imaging or machine learning contexts rather than in cervical cytology specifically and are rarely evaluated jointly within a single classification pipeline. No existing cervical cytology study selects a final model or ensemble using a composite criterion that spans discrimination, calibration, and selective prediction simultaneously, nor validates that selection's robustness to alternative metric weightings. The reliability-aware Hybrid-K ensemble selection framework proposed in this work directly addresses the core of this gap: candidate architectures are ranked and combined using an eight-metric composite score spanning all three reliability dimensions, and the resulting ensemble's robustness to metric weighting is explicitly quantified via Dirichlet-based sensitivity analysis. External validation on larger, independently collected, taxonomy-compatible single-cell cervical cytology datasets, which remains difficult given the taxonomy and image-format differences noted above, is identified as an important direction for future work rather than addressed here.

## 3. Methodology

Figure 2 provides an overview of the five-stage reliability-aware ensemble selection pipeline. The framework begins with dataset preparation and cross-validation design (Stage 1), followed by training nine candidate deep learning architectures under a shared experimental recipe (Stage 2). In Stage 3, each model is calibrated using temperature scaling and evaluated across eight classification and reliability metrics, which are combined into a composite ranking score. The top-ranked models are then assembled into Hybrid-K ensembles using soft voting (Stage 4). Finally, the selected ensemble is validated through Dirichlet-based robustness analysis, leave-one-metric-out ablation, and paired statistical testing (Stage 5). The subsections below describe each stage in detail.

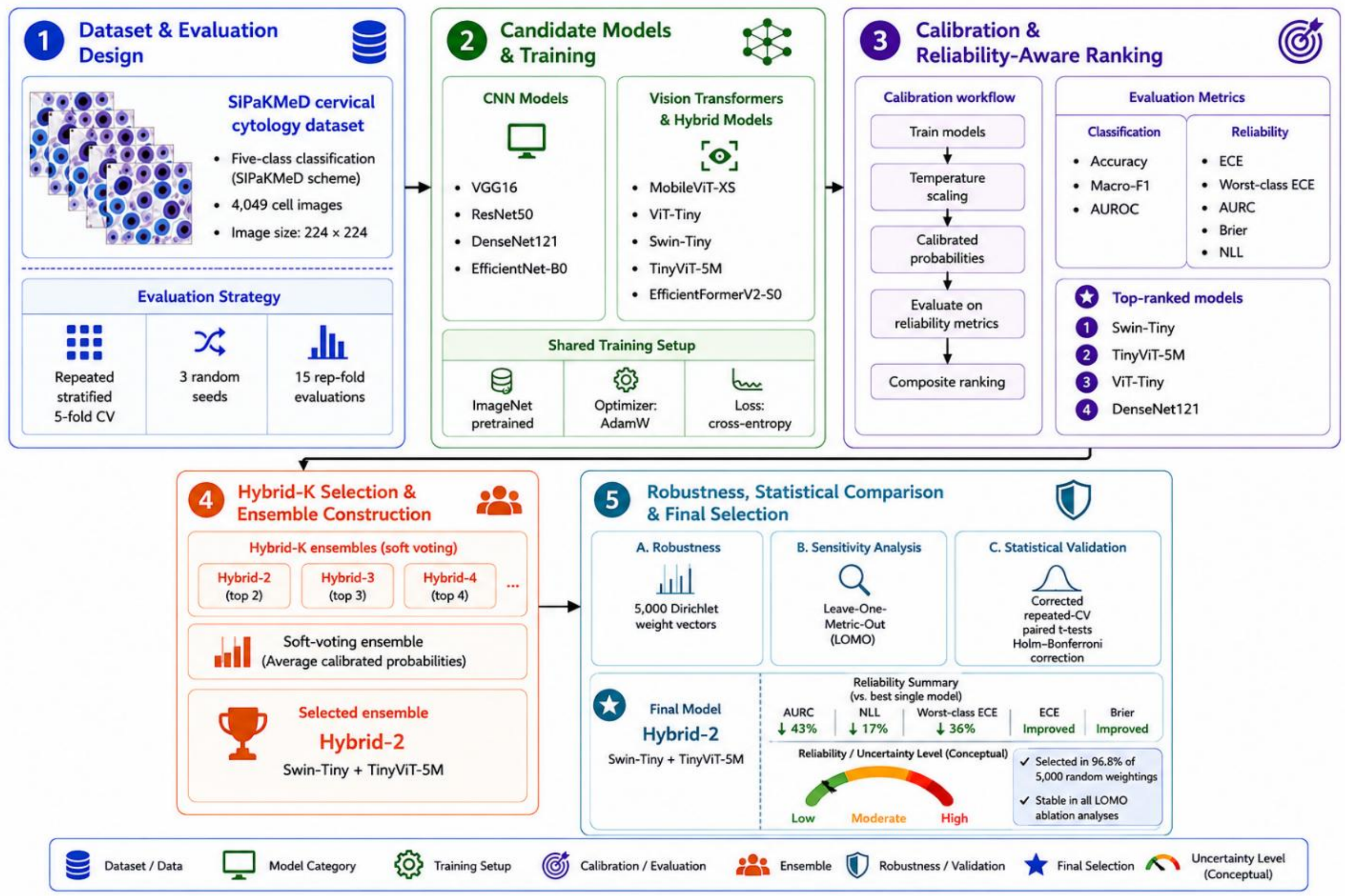

*Figure 2. Overview of the five-stage reliability-aware ensemble selection framework. Swin-Tiny and TinyViT-5M (Hybrid-2) were identified as the final ensemble across the internal evaluation stages on SIPaKMeD.*

### 3.1 Dataset and Cross-Validation Design

This study used the SIPaKMeD cervical cytology image dataset [4] for multiclass cervical cell classification. SIPaKMeD is a widely used public benchmark for cervical cytology classification that has been adopted by numerous prior deep learning studies [36, 65, 67-69], which makes it well suited for benchmarking the proposed framework against representative prior work (Section 4.10). SIPaKMeD was selected because it provides expert-annotated single-cell images spanning five cytomorphological classes with near-balanced class representation, and its public availability supports reproducibility and direct comparison with existing methods. The classification task was formulated as a five-class problem, in which each of the 4,049 single-cell images was assigned to one of five cytomorphological categories: Superficial-Intermediate, Parabasal, Koilocytotic, Dyskeratotic, and Metaplastic. As shown in Table 1, class representation is close to balanced (range: 787–831 images per class, 19.4%–20.5%), which limited the influence of class imbalance on model training and evaluation. The objective was not only to maximize predictive accuracy, but also to identify a model or ensemble configuration that provides reliable probability estimates, strong calibration, and improved selective-prediction behavior.

*Table 1. Class distribution of the SIPaKMeD dataset [4] used in this study.*

| Dataset | Class | N | % |
|---|---|---|---|
| **SIPaKMeD** | Superficial-Intermediate | 831 | 20.5% |
| | Parabasal | 787 | 19.4% |
| | Metaplastic | 793 | 19.6% |
| | Koilocytotic | 825 | 20.4% |
| | Dyskeratotic | 813 | 20.1% |
| | **Total** | **4,049** | **100.0%** |

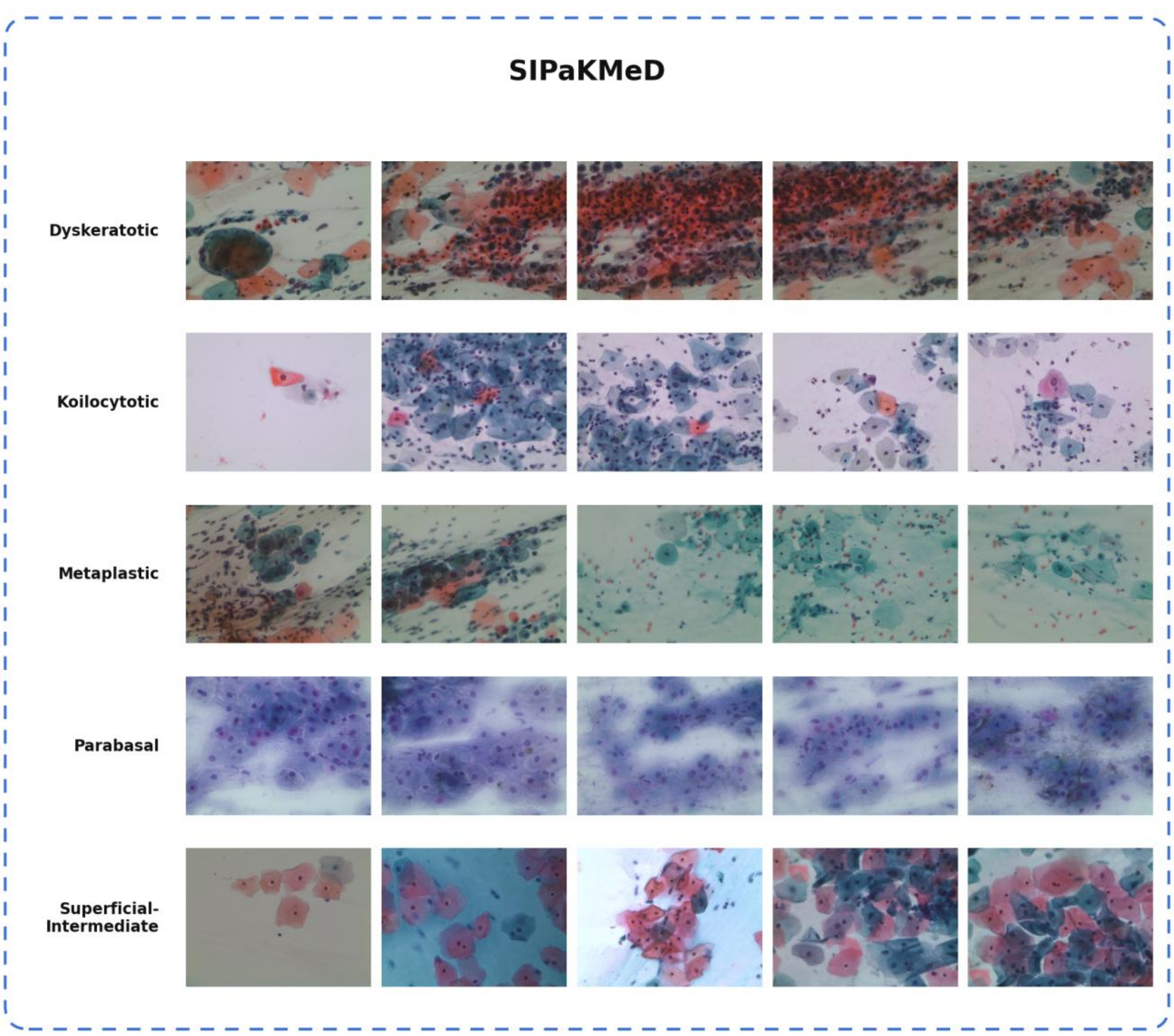

*Figure 3. Representative sample images from the SIPaKMeD dataset. Each row corresponds to one of the five cytomorphological classes used in this study (Dyskeratotic, Koilocytotic, Metaplastic, Parabasal, and Superficial-Intermediate) with five example images per class illustrating the within-class morphological and staining variability that the classification models must accommodate.*

To obtain a robust estimate of model performance, all experiments were conducted using stratified five-fold cross-validation. Stratification was used to preserve the class distribution across folds. Cross-validation was performed at the level of individual cropped single-cell images, which constitute the unit of analysis in this study. The SIPaKMeD cropped-cell filenames encode the parent cluster image from which each cell was segmented (for example, 001_01 and 001_02 share the parent identifier 001), so parent-image membership is recoverable from the released data; parent-level grouping, however, was not enforced in the stratified k-fold partitioning, which assigned cells independently on the basis of class label alone. The splitting procedure therefore permits cropped cells originating from the same parent image to be assigned to different folds. Accordingly, the results reported in this work characterize image-level internal validation and should not be interpreted as evidence of parent-image-, slide-, or specimen-level generalization. A single stratified five-fold partition was generated once with a fixed seed (42) and reused unchanged for every model, ensemble configuration, and replication. Model training was then repeated under three independent training seeds (42, 123, and 456), which varied weight initialization, augmentation, and sampler order but not the partition itself, producing a total of 15 fold-by-seed evaluations (5 fixed folds × 3 training seeds) for each model and ensemble configuration. These 15 evaluations therefore comprise five distinct held-out folds each evaluated under three training seeds, and are not three independently randomized cross-validation repetitions. The same train–test splits were used across all candidate models to ensure that performance comparisons were paired and fair. This design allowed the statistical analysis to compare models using matched fold-level scores rather than independent or unmatched estimates. This paired split design follows established practice for rigorous comparative benchmarking of machine learning, deep learning, and transformer models, where inconsistent train-test partitioning has been shown to inflate apparent performance differences [70].

### 3.2 Candidate Model Training

Nine deep learning models spanning three architecture families were evaluated as candidate classifiers. The families were conventional convolutional neural networks, vision transformers, and lightweight hybrid CNN–transformers. The candidate models were VGG16, ResNet50, DenseNet121, EfficientNet-B0, MobileViT-XS, ViT-Tiny, Swin-Tiny, TinyViT-5M, and EfficientFormerV2-S0. The model pool covered a broad architectural range, from classical CNN backbones to more recent transformer-based and hybrid architectures. The inclusion of ViT-Tiny and Swin-Tiny reflects growing evidence that vision transformer architectures can match or exceed CNN baselines on cervical cell classification when evaluated under statistically rigorous, reproducible benchmarking protocols [27-29, 71, 72].

All candidate models were trained using the same experimental recipe to ensure a fair comparison. Input images were resized to 224 × 224 pixels and normalized using ImageNet mean and standard deviation. ImageNet-pretrained weights were used for all backbones, and the final classification layer was replaced to match the five-class cervical cytology task. Each model was trained for 15 epochs with a batch size of 32. AdamW was used as the optimizer with a learning rate of $1 \times 10^{-4}$ and weight decay of $1 \times 10^{-4}$. A linear warmup of two epochs followed by cosine learning-rate decay was used to stabilize optimization. Cross-entropy loss with label smoothing of 0.1 was used during training. Although the SIPaKMeD classes are only mildly imbalanced, a weighted random sampler was applied at the data level to guarantee approximately class-balanced mini-batches across all folds and seeds, keeping the shared training recipe identical for every model rather than to correct a severe imbalance. Using an identical training recipe across all nine models ensured that observed differences in performance were attributable primarily to model architecture and reliability behavior rather than unequal hyperparameter tuning.

### 3.3 Calibration and Reliability Evaluation

Because high classification accuracy alone is insufficient in high-stakes medical image classification, each model was evaluated using classification and reliability-oriented metrics. Three discrimination metrics were used: macro-F1, accuracy, and AUROC. These metrics measured the ability of the models to correctly classify cervical cytology images across the five classes.

Five additional metrics were used to evaluate reliability, calibration, probabilistic quality, and selective prediction. ECE was used to measure the agreement between predicted confidence and empirical accuracy. WC-ECE was included to evaluate calibration behavior in the least-calibrated class, which is important in imbalanced or clinically sensitive classification problems. AURC was used to evaluate selective prediction, where a lower value indicates better ability to reduce classification risk when uncertain cases are rejected. Brier score and NLL were used to assess the quality of probabilistic predictions.

Post-hoc temperature scaling was applied to calibrate the predicted probabilities. Temperature scaling adjusts the confidence distribution without changing the predicted class labels. Calibration was implemented as a leave-one-fold-out procedure operating on the stored held-out predictions rather than on a separate calibration split. For each of the 15 fold-by-seed evaluations, a single scalar temperature was obtained by minimizing the negative log-likelihood over the pooled held-out predictions of the remaining 14 evaluations, and the fitted temperature was then applied to the target fold. Because the five-fold partition was fixed and only the training seed varied across replications, those remaining 14 evaluations include the two training-seed replicates of the target fold, which cover the same images and labels as the target fold itself; the data used to estimate the temperature for a given fold are therefore not fully independent of that fold. In addition, the scaled and unscaled probabilities of the target fold were compared and the variant with the lower expected calibration error on that fold was retained, so the decision of whether to apply the fitted temperature also depended on the target fold. No inner calibration split was withheld from the training data at any point. Calibration was evaluated after this step to determine whether each model provided reliable probability estimates in addition to accurate class predictions.

### 3.4 Reliability-Aware Composite Ranking

To select models using predictive performance and reliability, a reliability-aware composite score was computed from the eight evaluation metrics: macro-F1, accuracy, AUROC, ECE, WC-ECE, AURC, Brier score, and NLL. Because these metrics do not share a common direction (higher values are better for some and worse for others), all metrics were first transformed so that higher normalized values indicated better performance. For classification metrics (macro-F1, accuracy, and AUROC), higher values were already better. For reliability and loss-based metrics (ECE, WC-ECE, AURC, Brier score, and NLL), lower values were preferable; their normalized scores were therefore inverted before computing the composite.

Each metric was assigned equal weight ($w_1, \ldots, w_8 = 1/8$). Equal weighting was used to avoid privileging accuracy over reliability or introducing subjective group-level preferences. The final composite score was calculated as the average of the eight normalized metric scores. Candidate models were ranked according to this composite.

### 3.5 Hybrid-K Ensemble Construction

After ranking the candidate models, Hybrid-K ensembles were constructed using the top-ranked models from the reliability-aware composite ranking. Hybrid-2 was formed using the two highest-ranked models, Hybrid-3 using the three highest-ranked models, and Hybrid-4 using the four highest-ranked models. The purpose of evaluating multiple ensemble sizes was to determine whether adding more models produced meaningful gains beyond the smallest ensemble. A smaller ensemble was preferred when it captured most of the reliability benefit without adding unnecessary complexity.

To clarify the prediction pathway of the selected compact ensemble, Figure 4 illustrates the Hybrid-2 soft-voting mechanism used to combine the Swin-Tiny and TinyViT-5M probability outputs.

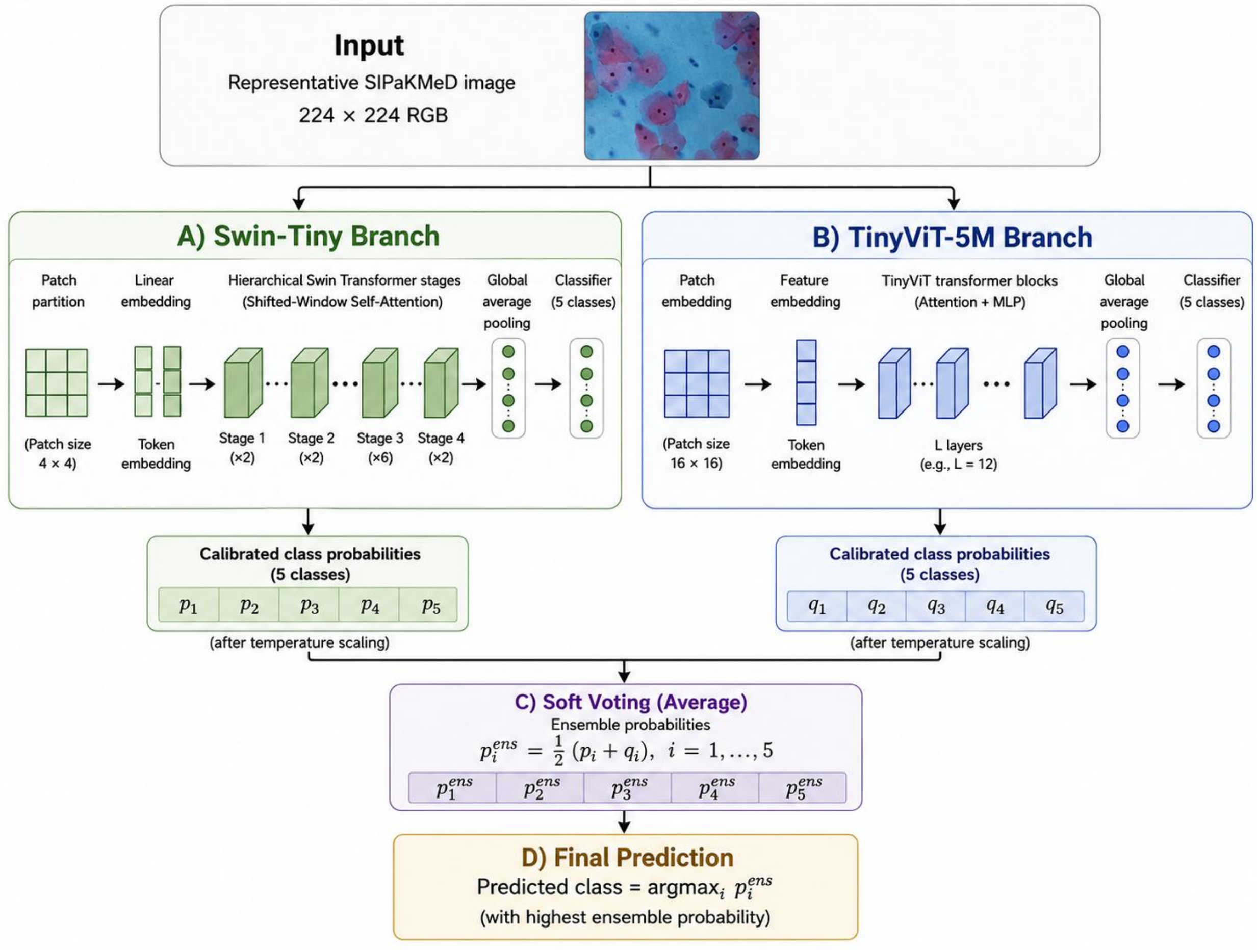

*Figure 4. Schematic of the selected internal Hybrid-2 soft-voting ensemble. The same representative SIPaKMeD input image is processed independently by the Swin-Tiny and TinyViT-5M branches to produce calibrated five-class probability vectors. The two vectors are averaged using unweighted soft voting, and the final prediction is assigned to the class with the highest ensemble probability.*

### 3.6 Robustness Analysis Using Random Metric Weightings

Although the main ranking used equal metric weights, a robustness analysis was performed to evaluate whether the selected ensembles were sensitive to the equal-weight assumption. The use of random weight perturbation to assess ranking stability is well-established in stochastic multicriteria acceptability analysis (SMAA), where Monte Carlo integration over weight distributions is the standard approach for evaluating decisions under incomplete preference information [59]. The Dirichlet distribution was specifically chosen because it generates weight vectors that sum to one, making it the natural distribution for exploring the full weight simplex in multicriteria frameworks [60].

To implement this analysis, 5,000 random metric-weight vectors were sampled from a symmetric Dirichlet distribution ($\alpha = 1$) over the eight metrics. Each sampled vector represented an alternative weighting scenario in which the relative importance of the eight metrics varied while the total weight remained equal to one. The main robustness sample size of N = 5,000 was selected based on the Monte Carlo sample-size analysis of Mazurek & Strzałka [61], who derive the analytical relationship $N = 1/(3p^2)$ between the required sample size and the relative standard error p. Their Table 10 shows that approximately N = 3,333 samples are sufficient to achieve a 1% relative standard error ($p = 0.01$); using N = 5,000 therefore exceeds this requirement, providing stable and precise estimates while maintaining reasonable computational cost. Their Figure 4 confirms empirical convergence of utility estimates by N = 5,000 across the tested range of N = 1,000–10,000. Additional sample sizes ranging from 500 to 10,000 were evaluated solely to verify convergence of selection frequencies and were not used to optimize any reported result.

For each sampled metric-weight vector, the composite score was recomputed, the nine candidate models were re-ranked, and the corresponding Hybrid-2, Hybrid-3, and Hybrid-4 ensembles were recorded. The modal selection frequency of each Hybrid-K configuration (that is, the fraction of random weighting scenarios in which a given configuration was the most-selected outcome) was then calculated across the 5,000 draws. A high modal selection frequency indicated that the selected ensemble was robust to changes in the relative importance assigned to individual metrics.

### 3.7 Leave-One-Metric-Out Sensitivity Analysis

To evaluate metric dependence, a leave-one-metric-out (LOMO) sensitivity analysis was conducted. Each of the eight metrics was removed once, and the composite ranking was recomputed using the remaining seven metrics with equal weights. After each metric was removed, Hybrid-2, Hybrid-3, and Hybrid-4 were reconstructed from the updated ranking and compared with the original full-metric selections.

A metric was considered selection-influential if removing it changed the membership of at least one Hybrid-K level (Hybrid-2, Hybrid-3, or Hybrid-4) compared with the full eight-metric baseline. A stable ensemble would remain unchanged even when any one metric was removed.

### 3.8 Statistical Comparison

Paired statistical comparisons were performed using the 15 fold-by-seed evaluations (5 fixed folds × 3 training seeds). The main statistical comparison evaluated the selected Hybrid-2 ensemble against the best individual model. Additional comparisons evaluated whether larger ensembles (Hybrid-3 and Hybrid-4) provided statistically meaningful improvement over Hybrid-2. For each metric, the corrected repeated cross-validation paired t-test of Nadeau and Bengio was computed on the 15 fold-by-seed differences; this corrected resampled test rescales the variance of the fold differences by $(1/n + 1/(k-1))$, with $k = 5$ folds and $n = 15$, to account for the correlation among fold scores induced by overlapping training sets in k-fold cross-validation, which the ordinary paired t-test ignores. Because the same five held-out folds were reused across the three training seeds, these 15 fold-by-seed observations (5 fixed folds × 3 training seeds) carry additional dependence that is not fully represented by the corrected resampled-test formulation. The reported p-values should therefore be interpreted as approximate. Within each pairwise comparison, the resulting eight metric-level p-values were adjusted for multiple comparisons using the Holm-Bonferroni procedure, and significance was assessed at $p < 0.05$ on the Holm-adjusted p-values. The purpose of this analysis was twofold: first, to determine whether the proposed ensemble significantly improved performance and reliability compared with the best individual classifier; and second, to determine whether larger ensembles provided sufficient additional benefit to justify their increased complexity.

The final ensemble was selected by considering metric performance, robustness under random weighting, LOMO stability, statistical significance, and model parsimony.

### 3.9 Implementation Details and Computational Environment

The experimental pipeline was implemented in Python using the PyTorch deep learning framework. Image preprocessing and augmentation were performed with torchvision, and candidate backbone architectures were implemented using timm version 1.0.11. Model training used the AdamW optimizer, automatic mixed precision where enabled, a two-epoch linear warmup followed by cosine-annealed learning-rate scheduling, cross-entropy loss with label smoothing, and class-balanced sampling through a weighted random sampler. Cross-validation splitting and evaluation metrics were computed with scikit-learn, numerical and tabular processing used NumPy and pandas, statistical comparisons used SciPy, and figures were generated with Matplotlib and seaborn.

Training and evaluation were conducted on Google Colab using an NVIDIA A100 GPU runtime with CUDA acceleration. The code included runtime checks for GPU availability, GPU device name, and total GPU memory at the start of each session. To support reproducibility, random seeds were fixed for Python, NumPy, and PyTorch, including CUDA; fold-level results were checkpointed to disk as each fold completed; and the same fixed stratified five-fold partition described in Sections 3.1–3.8 was used for every reported experiment.

### 3.10 Evaluation Protocol

This subsection consolidates the eight evaluation metrics used throughout the study and the equal-weight composite score used for reliability-aware model selection. Unless stated otherwise, all metrics are computed on the pooled, calibrated held-out predictions across the fixed stratified five-fold partition described in Sections 3.1–3.3. Let N denote the number of evaluation samples, C the number of classes, $y_i \in \{1, \ldots, C\}$ the true label of sample i, and $p_i = (p_{i1}, \ldots, p_{iC})$ the predicted class-probability vector with $\hat{y}_i = \text{argmax}_k \, p_{ik}$ denoting the predicted label. Arrows indicate whether higher (↑) or lower (↓) values are better.

**Discrimination metrics**

Accuracy (↑) is the fraction of correctly classified samples:

$$\text{Accuracy} = \frac{1}{N}\sum_{i=1}^{N} \mathbb{1}\,(\hat{y}_i = y_i)$$

Macro-F1 (↑) averages the per-class F1 scores, each combining the class-wise precision and recall:

$$\text{Macro} - \text{F1} = \frac{1}{C}\sum_{c=1}^{C}\frac{2\,\text{Prec}_c\,\text{Rec}_c}{\text{Prec}_c + \text{Rec}_c}$$

AUROC (↑) averages the per-class one-vs-rest areas under the receiver operating characteristic curve, obtained by treating class c as positive and the remaining classes as negative:

$$\text{AUROC} = \frac{1}{C}\sum_{c=1}^{C}\text{AUROC}_c$$

**Calibration and probabilistic-reliability metrics**

ECE (↓) partitions the N predictions into M = 15 equal-width confidence bins $B_m$ by their top-class confidence $\text{conf}(p_i) = \max_k p_{ik}$, and sums the weighted gap between per-bin accuracy $\text{acc}(B_m)$ and average confidence $\text{conf}(B_m)$:

$$\text{ECE} = \sum_{m=1}^{M}\frac{|B_m|}{N}|\text{acc}(B_m) - \text{conf}(B_m)|$$

WC-ECE (↓) is the maximum of the class-conditional ECE values, where the class-conditional ECE is computed exactly as above but restricted to samples whose true label is c:

$$\text{WC} - \text{ECE} = \max_c \text{ECE}_c$$

The multiclass Brier score (↓) is the mean squared error between the predicted probability vector and the one-hot encoded label vector $y_i$ (with entry $y_{ik} = 1$ if $k = y_i$ and 0 otherwise):

$$\text{Brier} = \frac{1}{N}\sum_{i=1}^{N}\sum_{k=1}^{C}(p_{ik} - y_{ik})^2$$

NLL (↓), equivalently the cross-entropy, penalizes the log-probability assigned to the true class:

$$\text{NLL} = -\frac{1}{N}\sum_{i=1}^{N}\log p_{i,y_i}$$

AURC (↓) summarizes selective-prediction behavior. Samples are ranked by descending confidence; at coverage τ (the fraction of most-confident samples retained), the selective risk R(τ) is the error rate on that retained subset, and AURC integrates risk over coverage:

$$\text{AURC} = \int_0^1 R\,(\tau)\,d\tau$$

**Reliability-aware composite score**

To rank the nine candidate models on a common scale, each of the eight metrics is min–max normalized across the model pool so that 1 denotes the best value and 0 the worst. For a higher-is-better metric with value $v_j$ for model j, the normalized value is:

$$\tilde{v}_j = \frac{v_j - \min v}{\max v - \min v}$$

For a lower-is-better metric (ECE, WC-ECE, Brier, NLL, AURC), the normalized value is inverted so that higher is always better:

$$\tilde{v}_j = 1 - \frac{v_j - \min v}{\max v - \min v}$$

The reliability-aware composite score is the unweighted mean of the eight normalized metrics, giving equal 1/8 weight to each; models are ranked in descending order of $S_j$, and this ranking drives the Hybrid-K selection in Section 3.5:

$$S_j = \frac{1}{8}\sum_{m=1}^{8} \tilde{v}_{j,m}$$

where ṽ(j,m) is the normalized value of metric m for model j. The three discrimination metrics (accuracy, macro-F1, AUROC) and five reliability metrics (ECE, WC-ECE, Brier, NLL, AURC) thus contribute equally to the default composite score; alternative weightings are examined in the robustness and sensitivity analyses of Sections 3.6 and 3.7.

## 4. Results and Discussion

This section reports the empirical findings of the reliability-aware selection framework on the SIPaKMeD five-class cervical cytology task. The results are organized to follow the methodology: Section 4.1 describes the experimental design and evaluation setting; Section 4.2 reports per-model performance and ranking for the nine candidate architectures; Section 4.3 presents calibration and reliability results; and Section 4.4 derives the composite ranking used for Hybrid-K selection. Section 4.5 evaluates the resulting Hybrid-K ensembles, while Sections 4.6 and 4.7 assess robustness to random metric weightings and leave-one-metric-out sensitivity, respectively. Section 4.8 reports the paired statistical comparisons, and Section 4.9 states the final ensemble selection. Section 4.10 then situates the framework among representative SIPaKMeD studies.

### 4.1 Experimental Design and Evaluation Setting

All models were evaluated under the same fixed stratified five-fold partition with three training seeds described in the methodology. Training was then repeated under three independent training seeds while holding this partition fixed, producing 15 fold-by-seed evaluations (5 fixed folds × 3 training seeds) for each candidate model and ensemble configuration. The same splits were used across all models, enabling paired comparisons between architectures, between individual models and ensembles, and between different ensemble sizes. This evaluation setting provided fold-level estimates for classification metrics, calibration metrics, probabilistic reliability metrics, and selective-prediction metrics, which were also used for the paired statistical tests reported in Section 4.8.

### 4.2 Candidate Model Performance and Ranking

The nine candidate models were first evaluated individually using the eight-metric reliability-aware framework. Swin-Tiny achieved the highest composite score (0.939) and was identified as the best individual model. TinyViT-5M ranked second (0.876), followed by ViT-Tiny (0.787) and DenseNet121 (0.781). The results show that the top-ranked models combined high classification performance with favorable reliability behavior. ViT-Tiny's competitive third-place ranking is consistent with prior reproducible benchmarking showing that lightweight vision transformers can match established CNN baselines on cervical cell classification. Based on this ranking, Swin-Tiny and TinyViT-5M were selected as the two members of the Hybrid-2 ensemble. Table 2 reports the full per-metric scores and composite rankings for all nine candidate models.

Across the nine candidate models, discrimination was tightly clustered (macro-F1 ranged from 0.912 to 0.982 and AUROC from 0.991 to 0.999), so top-line accuracy alone did not separate the architectures. The composite ranking was instead driven by the reliability-oriented metrics (Brier score, NLL, AURC, and ECE), as intended by the reliability-aware framework. Consistent with the motivation for including lightweight vision transformers, the three transformer models (Swin-Tiny, TinyViT-5M, and ViT-Tiny) occupied the top three ranks, with the strongest convolutional baseline, DenseNet121, close behind in fourth and the remaining convolutional and hybrid networks (MobileViT-XS, VGG16, and ResNet50) forming a competitive mid-pack. These outcomes were largely expected: Swin-Tiny led on nearly every metric, and DenseNet121's standing is consistent with the calibration and feature-reuse benefits often attributed to dense connectivity.

Two results were less expected. EfficientNet-B0 was a clear outlier, with the lowest composite score (0.138) driven not by a collapse in discrimination but by substantially weaker probabilistic reliability; its Brier score (0.129) and NLL (0.255) were several times those of the top-ranked models, suggesting that its compound-scaled representation was more poorly calibrated and less reliable under the shared training and temperature-scaling protocol. EffFormerV2-S0 showed a split profile, achieving the best worst-class ECE (0.022) yet the weakest selective-prediction behavior in the field (AURC 0.018, several times the best models' 0.003) together with an elevated NLL (0.147); strong worst-class calibration therefore did not translate into reliable confidence ranking. These cases reinforce the value of assessing reliability with multiple complementary metrics rather than accuracy alone, and they support the selection of Swin-Tiny and TinyViT-5M, which combined top discrimination with the most favorable reliability behavior.

***Table 2. Per-metric ranking of nine candidate models under the equal-weight reliability-aware composite score ($w_1 \dots w_8 = 1/8$). Bold rows indicate the Hybrid-2 members.***

| Rank | Model | F1 ↑ | Acc ↑ | AUROC ↑ | ECE ↓ | WC-ECE ↓ | AURC ↓ | Brier ↓ | NLL ↓ | Score ↑ |
|---|---|---|---|---|---|---|---|---|---|---|
| **1** | **Swin-Tiny** | **0.982** | **0.982** | **0.999** | **0.004** | **0.029** | **0.003** | **0.029** | **0.068** | **0.939** |
| **2** | **TinyViT-5M** | **0.977** | **0.977** | **0.999** | **0.006** | **0.027** | **0.003** | **0.037** | **0.079** | **0.876** |
| 3 | ViT-Tiny | 0.976 | 0.976 | 0.998 | 0.006 | 0.033 | 0.004 | 0.038 | 0.084 | 0.787 |
| 4 | DenseNet121 | 0.973 | 0.973 | 0.998 | 0.005 | 0.035 | 0.004 | 0.043 | 0.088 | 0.781 |
| 5 | MobileViT-XS | 0.969 | 0.969 | 0.998 | 0.008 | 0.027 | 0.004 | 0.050 | 0.105 | 0.744 |
| 6 | VGG16 | 0.971 | 0.971 | 0.998 | 0.008 | 0.035 | 0.003 | 0.045 | 0.096 | 0.708 |
| 7 | ResNet50 | 0.968 | 0.968 | 0.997 | 0.005 | 0.037 | 0.005 | 0.051 | 0.104 | 0.703 |
| 8 | EffFormerV2-S0 | 0.968 | 0.968 | 0.995 | 0.010 | 0.022 | 0.018 | 0.049 | 0.147 | 0.559 |
| 9 | EfficientNet-B0 | 0.912 | 0.913 | 0.991 | 0.005 | 0.036 | 0.015 | 0.129 | 0.255 | 0.138 |

***Note. WC-ECE = worst-class expected calibration error; AURC = area under the risk–coverage curve; NLL = negative log-likelihood. ↑ = higher is better; ↓ = lower is better. Score = eight-metric composite. Metrics evaluated after post-hoc temperature scaling. Bold rows (Swin-Tiny, TinyViT-5M) form the selected Hybrid-2 ensemble.***

### 4.3 Calibration and Reliability Results

After temperature scaling, the candidate models showed strong calibration behavior, with low ECE values across the model pool. The best individual model, Swin-Tiny, achieved an ECE of 0.0040 and WC-ECE of 0.0290. The Hybrid-2 ensemble preserved strong overall calibration (ECE = 0.0039) while improving worst-class calibration substantially (WC-ECE = 0.0186, a 36% reduction). Because the temperature parameter was estimated from held-out predictions that were not fully independent of the fold being evaluated, these calibration figures should be read as internal exploratory estimates of relative calibration behavior rather than as independently validated calibration performance.

Reliability gains were especially clear in selective prediction and probabilistic scoring. Compared with Swin-Tiny alone, Hybrid-2 reduced AURC from 0.0028 to 0.0016 (43% reduction), Brier score from 0.0292 to 0.0258, and NLL from 0.0679 to 0.0563 (17% reduction). These point-estimate improvements are consistent with Hybrid-2 providing more reliable confidence estimates and better risk-coverage behavior than the best individual classifier, subject to the calibration caveat noted above. The reduction in AURC is particularly important because it reflects that Hybrid-2 maintained lower classification risk as uncertain predictions were removed, supporting its potential value for reliability-aware selective prediction under the present internal-validation setting.

Across the nine individual models, the calibration results followed an expected pattern in aggregate but revealed informative differences in the tails. Post-hoc temperature scaling moved every model's reliability curve toward the diagonal (Figure 5) and produced uniformly low overall ECE (0.004 to 0.010), so aggregate calibration did not meaningfully separate the architectures. The more discriminating signals came from worst-class calibration and selective prediction: worst-class ECE ranged more widely (from 0.022 for EffFormerV2-S0 to 0.037 for ResNet50), and AURC clustered tightly for most models (0.003 to 0.005) except for EffFormerV2-S0 (0.018) and EfficientNet-B0 (0.015). These same two models also carried the highest NLL values (0.147 and 0.255), indicating that a low aggregate ECE can coexist with poor tail calibration and unreliable confidence ranking. This pattern is the main reason the composite score weighted worst-class ECE, AURC, and NLL alongside overall ECE, and it explains why models with near-identical accuracy separated clearly once probabilistic reliability was considered.

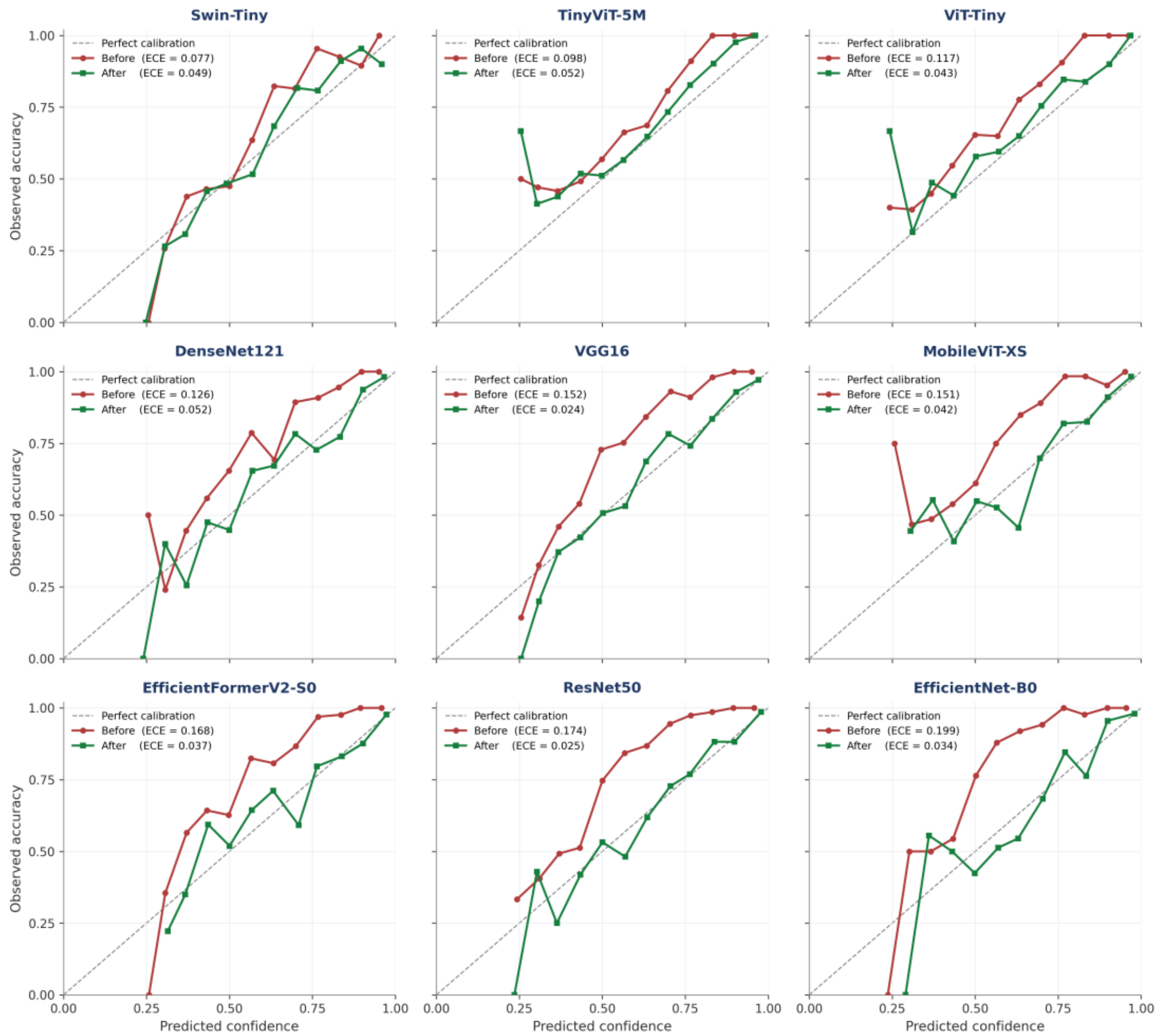

*Figure 5. Reliability diagrams for all nine candidate models before and after post-hoc temperature scaling. Temperature scaling moves every model's calibration curve toward the diagonal, improving ECE for all nine architectures.*

## 4.4 Composite Ranking and Hybrid-K Selection

Using the equal-weight eight-metric composite score, the top four models were Swin-Tiny, TinyViT-5M, ViT-Tiny, and DenseNet121. The reliability-aware ranking therefore produced the following ensemble configurations: Hybrid-2 consisted of Swin-Tiny and TinyViT-5M; Hybrid-3 added ViT-Tiny; and Hybrid-4 added DenseNet121. This confirms that ensemble members were selected based on a combined classification and reliability profile rather than accuracy alone.

## 4.5 Hybrid-K Ensemble Performance

The Hybrid-K ensembles improved several reliability metrics compared with the best individual model. Table 3 presents the full metric results for the best individual model and all three ensemble configurations.

***Table 3. Performance of the best individual model and Hybrid-K ensembles across all eight evaluation metrics. Bold row (Hybrid-2) indicates the selected final ensemble.***

| **Method** | **F1 ↑** | **Acc ↑** | **AUROC ↑** | **ECE ↓** | **WC-ECE ↓** | **AURC ↓** | **Brier ↓** | **NLL ↓** |
|---|---|---|---|---|---|---|---|---|
| Best individual (Swin-Tiny) | 0.9824 | 0.9824 | 0.9988 | 0.0040 | 0.0290 | 0.0028 | 0.0292 | 0.0679 |
| **Hybrid-2 (+ TinyViT-5M)** | **0.9847** | **0.9846** | **0.9992** | **0.0039** | **0.0186** | **0.0016** | **0.0258** | **0.0563** |
| Hybrid-3 (+ ViT-Tiny) | 0.9847 | 0.9847 | 0.9993 | 0.0037 | 0.0189 | 0.0013 | 0.0251 | 0.0529 |
| Hybrid-4 (+ DenseNet121) | 0.9843 | 0.9843 | 0.9992 | 0.0032 | 0.0206 | 0.0013 | 0.0253 | 0.0514 |

***Note. All values are means over 15 fold-by-seed evaluations (5 fixed folds × 3 training seeds). Hybrid-2 vs. best individual: AURC −43% (0.0028 → 0.0016), NLL −17% (0.0679 → 0.0563), WC-ECE −36% (0.0290 → 0.0186). Bold row = selected final ensemble.***

Relative to the best individual model, Hybrid-2 reduced AURC by approximately 43%, NLL by approximately 17%, and WC-ECE by approximately 36%. Although Hybrid-3 and Hybrid-4 produced small additional improvements for selected reliability metrics, the gains were limited compared with the improvement already achieved by Hybrid-2. Classification metrics were already near the ceiling for all configurations, so most of the meaningful improvement occurred in reliability-oriented metrics (AURC, Brier, NLL, WC-ECE). These findings suggest that Hybrid-2 captured most of the reliability benefit while maintaining the smallest ensemble size.

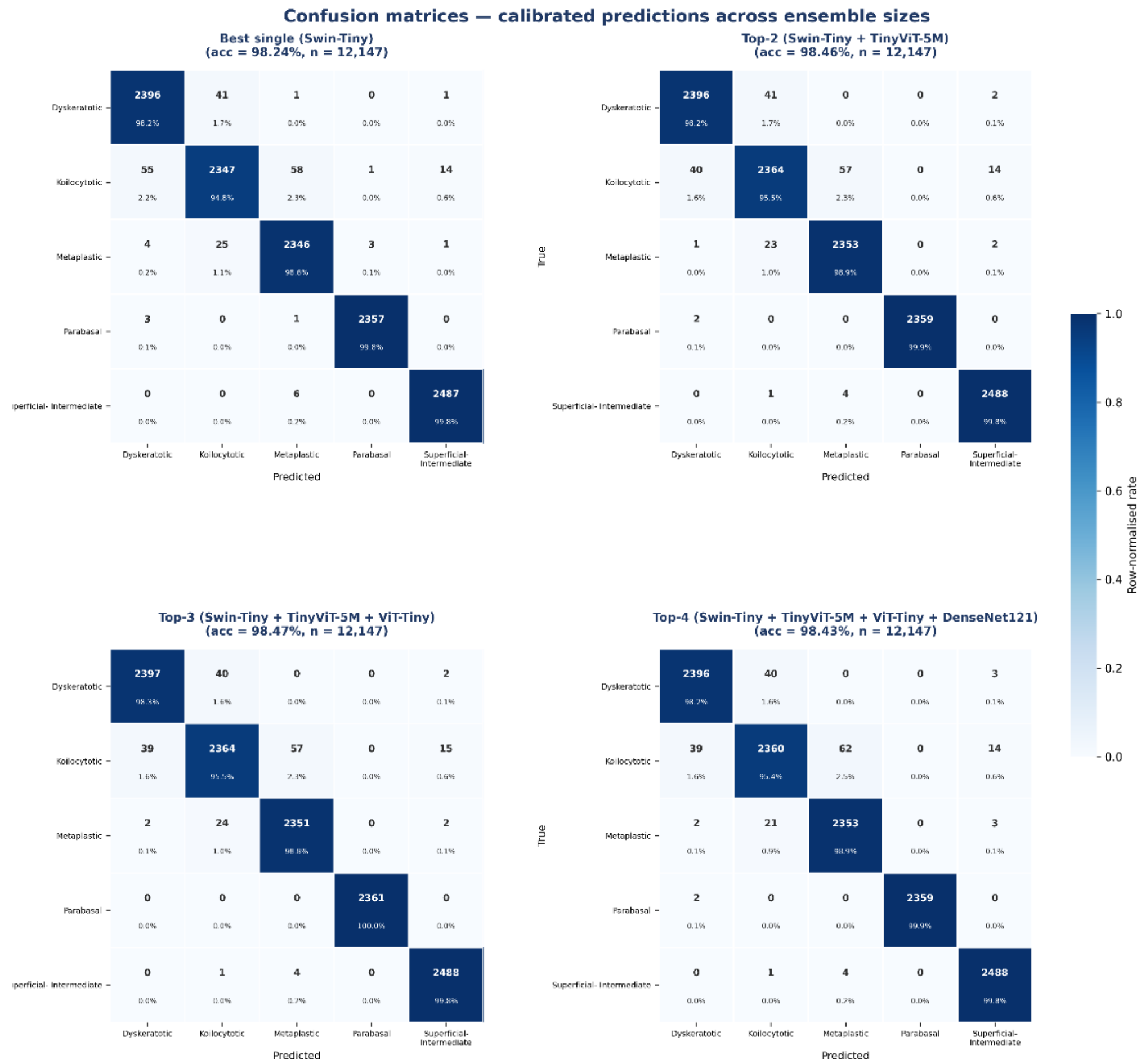

*Figure 6. Confusion matrices for the best individual model (Swin-Tiny) and Hybrid-2, Hybrid-3, and Hybrid-4 ensembles on the SIPaKMeD five-class task, using calibrated predictions. Hybrid-2 improves Koilocytotic recall (94.8% to 95.5%) and reduces most off-diagonal errors relative to the best individual model; gains taper for Hybrid-3 and Hybrid-4.*

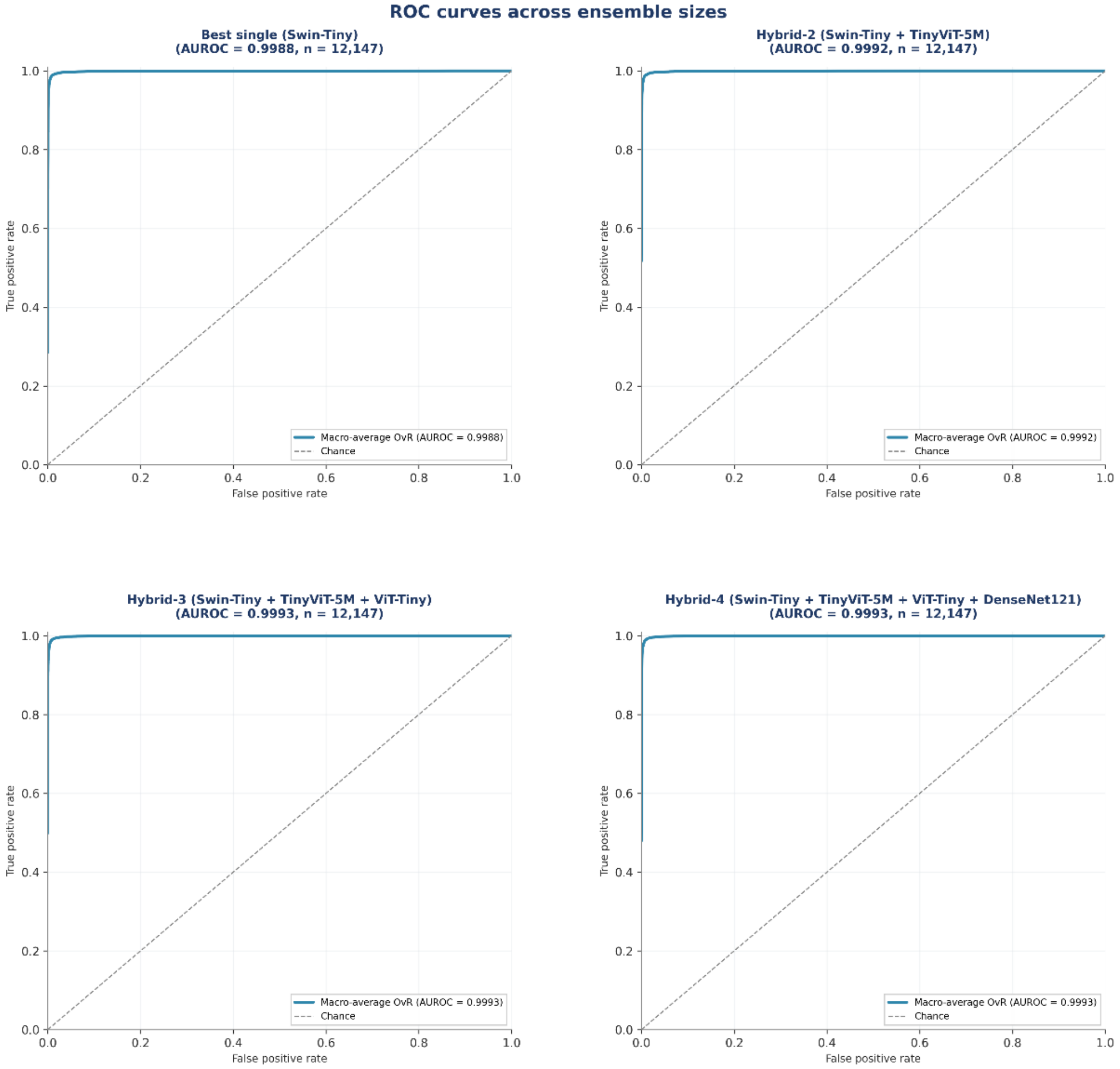

*Figure 7. Macro-average one-vs-rest ROC curves for the best individual model (Swin-Tiny) and the Hybrid-K ensembles on the SIPaKMeD five-class task. Each panel shows one model configuration separately for consistency with the confusion-matrix comparison in Figure 6. All configurations exhibit near-ceiling discrimination (AUROC = 0.9988–0.9993), with AUROC values remaining highly similar across ensemble sizes; the reliability-oriented metrics in Table 3 and Figure 9 provide the basis for selecting the compact Hybrid-2 ensemble.*

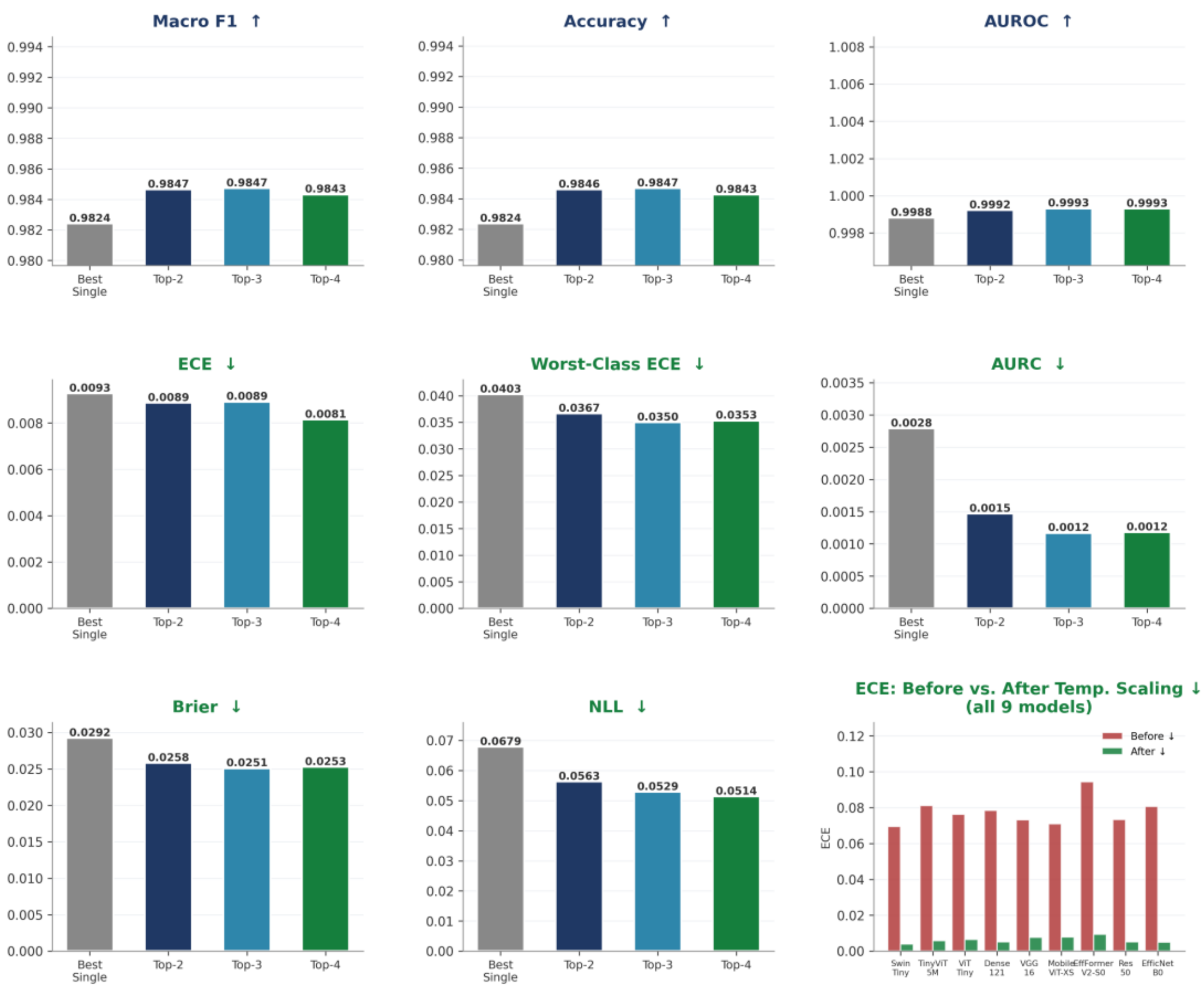

*Figure 8. Best individual model versus Hybrid-K ensembles across all eight evaluation metrics. Gains concentrate in the reliability metrics (AURC, Brier, NLL); classification metrics are already near ceiling for all configurations.*

The label "best single model" shown in the figure panels denotes the best individual model (Swin-Tiny); the two terms are equivalent and are used interchangeably here for consistency with the surrounding text.

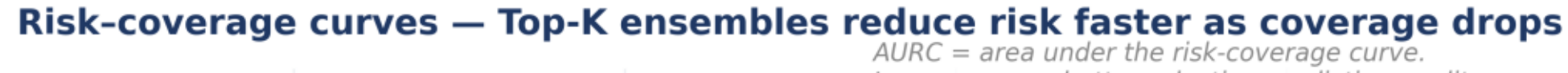

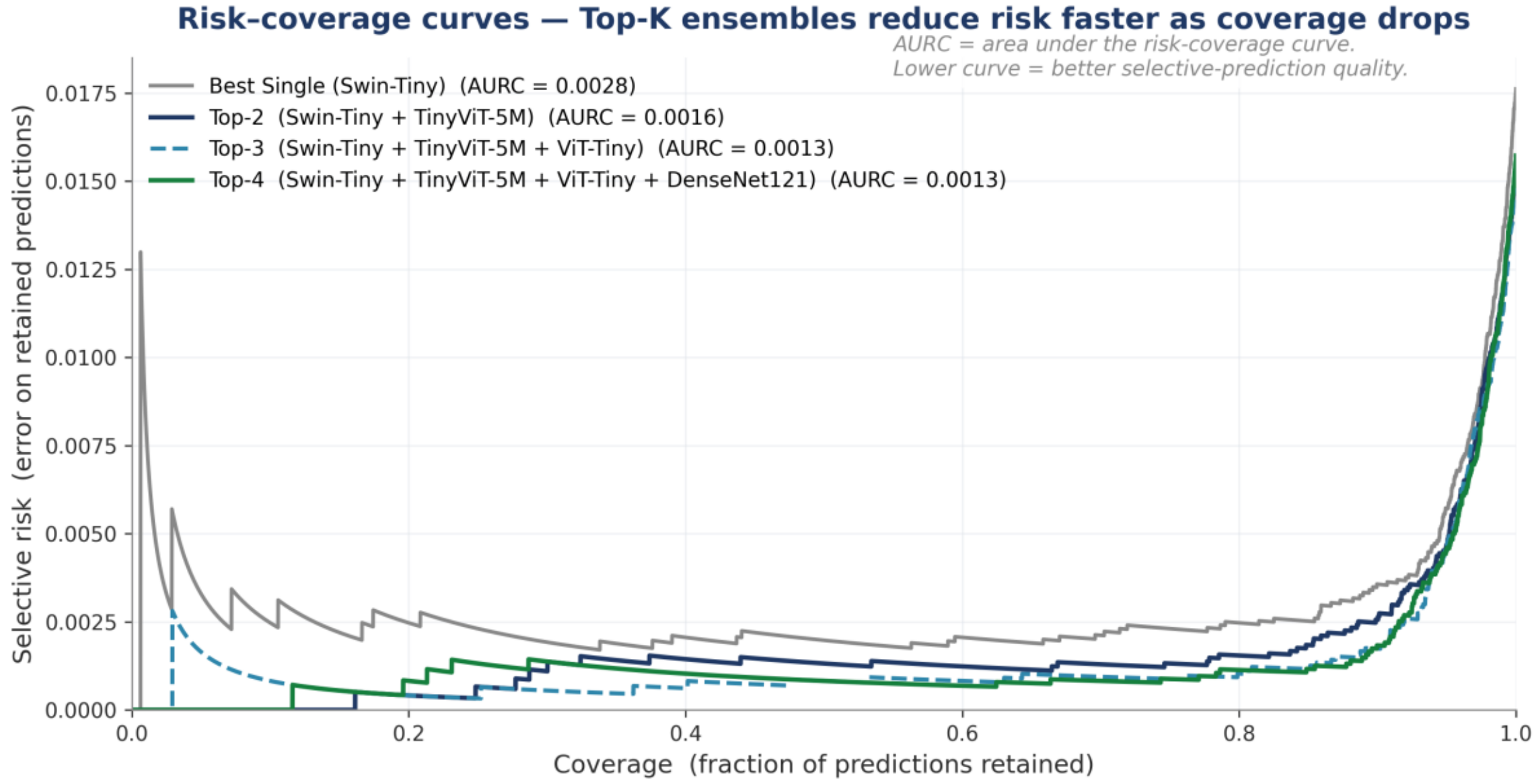

*Figure 9. Risk-coverage curves for Hybrid-K ensembles versus the best individual model. Hybrid-K ensembles maintain lower risk as coverage decreases, reflecting improved selective-prediction quality (lower AURC).*

## 4.6 Robustness Under Random Metric Weightings

The Dirichlet-based robustness analysis showed that the Hybrid-2 selection was highly stable under alternative metric-weighting scenarios. Table 4 reports the modal selection frequency of each Hybrid-K configuration across nested sample sizes from 500 to 10,000 Dirichlet draws.

***Table 4. Modal Hybrid-K selection frequency under nested prefixes of one Dirichlet stream (500 to 10,000 draws). The pre-specified main robustness sample size is N = 5,000 (bold row).***

| Dirichlet Draws (N) | Hybrid-2 Freq. (%) | Hybrid-3 Freq. (%) | Hybrid-4 Freq. (%) |
|---|---|---|---|
| 500 | 97.4% | 44.8% | 65.8% |
| 1,000 | 97.2% | 45.3% | 66.5% |
| 2,000 | 96.7% | 44.8% | 66.5% |
| **5,000 (main)** | **96.8%** | **45.0%** | **66.6%** |
| 10,000 | 96.8% | 45.2% | 67.2% |

***Note.*** Modal selection frequency = fraction of random weighting scenarios in which a given Hybrid-K configuration was the most-selected outcome. The N = 5,000 row was pre-specified as the main robustness estimate; remaining sample sizes verify convergence only and were not used to optimize any reported result.

Across the main set of 5,000 randomly sampled weight vectors, Hybrid-2 was selected in 96.8% of cases, indicating that the selection of Swin-Tiny and TinyViT-5M was not dependent on the equal-weight assumption. The modal selection frequency was stable across different numbers of Dirichlet draws (ranging from 96.7% to 97.4%), confirming that the robustness result did not depend on a particular number of random weighting samples. By comparison, Hybrid-3 (45.0%) and Hybrid-4 (66.6%) showed considerably lower and more variable modal selection frequencies, indicating greater sensitivity to changes in metric weighting. The higher consistency of Hybrid-4 relative to Hybrid-3 follows from their membership dynamics: ViT-Tiny and DenseNet121 are closely ranked at the third and fourth positions, so the third slot alternates between them under reweighting and leaves Hybrid-3's membership unstable, whereas Hybrid-4 contains both models and is therefore invariant to that swap. Therefore, Hybrid-2 was the most stable and robust ensemble configuration under alternative weighting assumptions.

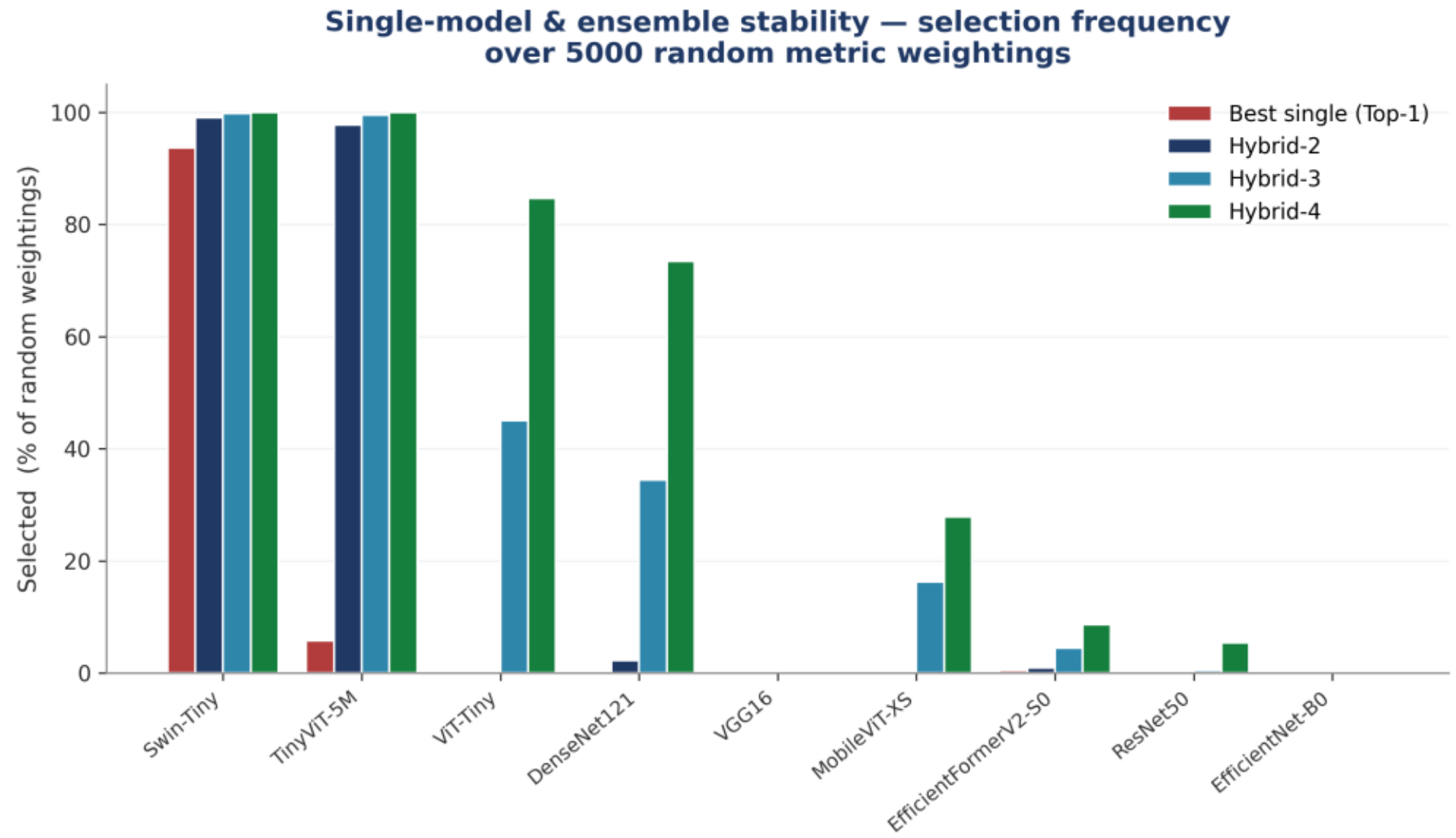

*Figure 10. Selection stability across 5,000 random Dirichlet-sampled metric weightings. Hybrid-2 (Swin-Tiny + TinyViT-5M) is selected as the modal ensemble in approximately 96.8% of weighting scenarios, and the best individual model in approximately 93.7%, indicating that the preferred configuration is robust to how the eight metrics are weighted.*

## 4.7 Leave-One-Metric-Out Sensitivity Results

The LOMO analysis confirmed the stability of the recommended Hybrid-2 ensemble. Table 5 presents the full results, showing the Hybrid-K membership after each of the eight metrics was removed one at a time.

***Table 5. Leave-one-metric-out (LOMO) ablation results. Each metric is removed once, and Hybrid-K selection is recomputed using the remaining seven metrics with equal weights. A metric is selection-influential if its removal changes the membership of at least one Hybrid-K level (Hybrid-2 [H2], Hybrid-3 [H3], or Hybrid-4 [H4]) relative to the full eight-metric baseline.***

| Dropped Metric | Group | H2 After Drop | H3 After Drop | H4 After Drop | Changed? | What Changed | Affects Any H-K? |
|---|---|---|---|---|---|---|---|
| **Macro-F1** | Classification | **Swin+TViT** | Swin+TViT+DN | Swin+TViT+DN+ViT | **Yes** | H3: +DenseNet; −ViT<br>H4: same swap | **Yes** |
| **Accuracy** | Classification | **Swin+TViT** | Swin+TViT+DN | Swin+TViT+DN+ViT | **Yes** | H3: +DenseNet; −ViT<br>H4: same swap | **Yes** |
| **AUROC** | Classification | **Swin+TViT** | Swin+TViT+ViT | Swin+TViT+ViT+DN | No | — | No |
| **ECE** | Reliability | **Swin+TViT** | Swin+TViT+ViT | Swin+TViT+ViT+MViT | **Yes** | H4: +MobileViT; −DenseNet | **Yes** |

| Dropped Metric | Group | H2 After Drop | H3 After Drop | H4 After Drop | Changed? | What Changed | Affects Any H-K? |
|---|---|---|---|---|---|---|---|
| **WC-ECE** | Reliability | **Swin+TViT** | Swin+TViT+DN | Swin+TViT+DN+ViT | **Yes** | H3: +DenseNet; −ViT<br>H4: same swap | **Yes** |
| **AURC** | Reliability | **Swin+TViT** | Swin+TViT+ViT | Swin+TViT+ViT+DN | No | — | No |
| **Brier** | Reliability | **Swin+TViT** | Swin+TViT+DN | Swin+TViT+DN+ViT | **Yes** | H3: +DenseNet; −ViT<br>H4: same swap | **Yes** |
| **NLL** | Reliability | **Swin+TViT** | Swin+TViT+ViT | Swin+TViT+ViT+DN | No | — | No |

***Note. Swin = Swin-Tiny; TViT = TinyViT-5M; ViT = ViT-Tiny; DN = DenseNet121; MViT = MobileViT-XS. H2 After Drop values are bolded to indicate that Hybrid-2 membership remains unchanged in all cases. WC-ECE = worst-class ECE.***

When each of the eight metrics was removed one at a time, Hybrid-2 remained unchanged in every case: Swin-Tiny and TinyViT-5M were consistently selected as the top two models regardless of which metric was excluded. Changes were observed only for larger ensembles. Removing macro-F1, accuracy, WC-ECE, or Brier score changed the membership of Hybrid-3 and Hybrid-4 (Hybrid-3 swapped ViT-Tiny for DenseNet121; Hybrid-4 underwent the corresponding shift). Removing ECE changed the membership of Hybrid-4 only (MobileViT-XS replaced DenseNet121), while Hybrid-3 remained unaffected. Removing AUROC, AURC, or NLL left all three Hybrid-K levels unchanged.

These findings confirm that the final recommended Hybrid-2 ensemble was not driven by any single evaluation metric. In contrast, Hybrid-3 and Hybrid-4 were more sensitive to metric composition, further supporting the conclusion that larger ensembles are less stable than the two-model configuration.

## 4.8 Statistical Comparison Results

Pairwise model comparisons used the corrected repeated cross-validation paired t-test (Nadeau and Bengio) across the 15 fold-by-seed evaluations (5 fixed folds × 3 training seeds), with Holm-Bonferroni correction applied across the eight metric-level p-values within each comparison. This corrected test accounts for the correlation among fold-level scores induced by overlapping training sets in k-fold cross-validation, which the ordinary paired t-test ignores. Table 6 reports the raw and Holm-adjusted p-values for all pairwise comparisons.

***Table 6. Corrected repeated CV paired t-test p-values with Holm-Bonferroni correction, across 15 fold-by-seed evaluations (5 fixed folds × 3 training seeds). The corrected resampled t-test (Nadeau and Bengio) rescales the variance of the fold differences by $(1/n + 1/(k-1))$ to account for the overlap between training sets across folds; Holm-Bonferroni correction is applied across the eight metric p-values within each comparison. Each cell reports raw two-sided / Holm-adjusted p-values. No comparison reached significance at $p < 0.05$ after correction.***

| Comparison | F1 | Acc | AUROC | ECE | WC-ECE | AURC | Brier | NLL |
|---|---|---|---|---|---|---|---|---|
| **Hybrid-2 vs. Best individual (Swin-Tiny)** | 0.1406 / 0.8434 | 0.1460 / 0.8434 | 0.2314 / 0.9256 | 0.8318 / 0.9256 | 0.4207 / 0.9256 | 0.2923 / 0.9256 | 0.0779 / 0.5450 | 0.0211 / 0.1684 |
| **Hybrid-2 vs. Hybrid-3** | 0.9490 / 1.0000 | 0.9486 / 1.0000 | 0.5067 / 1.0000 | 0.9629 / 1.0000 | 0.6716 / 1.0000 | 0.5537 / 1.0000 | 0.5410 / 1.0000 | 0.1957 / 1.0000 |
| **Hybrid-2 vs. Hybrid-4** | 0.7525 / 1.0000 | 0.7543 / 1.0000 | 0.4284 / 1.0000 | 0.5666 / 1.0000 | 0.7145 / 1.0000 | 0.5359 / 1.0000 | 0.5965 / 1.0000 | 0.0538 / 0.4305 |
| **Hybrid-3 vs. Hybrid-4** | 0.6507 / 1.0000 | 0.6535 / 1.0000 | 0.9177 / 1.0000 | 0.5120 / 1.0000 | 0.9226 / 1.0000 | 0.9716 / 1.0000 | 0.7991 / 1.0000 | 0.4358 / 1.0000 |

***Note.*** Best individual refers to Swin-Tiny. n = 15 paired fold-by-seed observations per comparison (5 fixed folds × 3 training seeds). Each cell shows raw two-sided p-value / Holm-adjusted p-value from the corrected repeated cross-validation paired t-test (Nadeau and Bengio); Holm-Bonferroni correction is applied across the eight metrics within each comparison. Because this table

reports p-values only, the direction and magnitude of differences should be interpreted from the metric means in Table 3. No comparison is significant at $p < 0.05$ after correction.

Non-significance should not be interpreted as evidence of no practical reliability gain; rather, it indicates that the observed per-metric improvements were not statistically distinguishable from fold-by-seed variability after multiple-comparison correction. Therefore, Hybrid-2 is selected based on the combined evidence of improved reliability point estimates, higher composite score, robustness to metric weighting, LOMO stability, and parsimony.

Compared with the best individual model (Swin-Tiny), Hybrid-2 improved the point estimates of most metrics, for example, reducing AURC from 0.0028 to 0.0016, Brier from 0.0292 to 0.0258, and NLL from 0.0679 to 0.0563, and raising macro-F1 from 0.9824 to 0.9847. Under the corrected repeated cross-validation t-test, the smallest raw p-value was obtained for NLL ($p = 0.0211$), followed by Brier score ($p = 0.0779$); macro-F1 ($p = 0.1406$), accuracy ($p = 0.1460$), AUROC ($p = 0.2314$), AURC ($p = 0.2923$), worst-class ECE ($p = 0.4207$), and ECE ($p = 0.8318$) had larger raw p-values. After Holm-Bonferroni correction across the eight metrics, none of these differences reached significance at $p < 0.05$ (all Holm-adjusted $p \geq 0.168$). The corrected analysis therefore does not provide statistical evidence that Hybrid-2 outperforms the best individual model on any individual metric; the improvements it offers are consistent in direction but fall within the variability expected across correlated fold-by-seed evaluations.

Despite the lack of statistical significance after correction, the standardized effect sizes (Cohen's $d_z$, computed from the same paired fold-by-seed differences underlying the corrected test) indicate that most of these improvements are of medium-to-large practical magnitude, all favoring Hybrid-2: $|d_z| = 1.46$ for NLL, 1.07 for Brier score, 0.88 for macro-F1, 0.87 for accuracy, 0.70 for AUROC, and 0.62 for AURC. Worst-class ECE showed a small effect ($|d_z| = 0.47$), and ECE showed a negligible effect ($|d_z| = 0.12$). These effect sizes are reported as descriptive, exploratory quantities rather than as independent estimates of effect magnitude: the three training-seed evaluations within each fold share the same held-out samples, so the 15 paired differences are not independent observations and the standardized magnitudes should be read as summaries of the observed differences rather than as inferential quantities. Under conventional benchmarks (Cohen, 1988), effect sizes above 0.8 are considered large and those between 0.5 and 0.8 medium; the observed improvements are therefore consistent and non-trivial in magnitude, even though most of them do not reach significance after multiple-comparison correction under the corrected test with $n = 15$ correlated fold-by-seed observations.

When Hybrid-2 was compared with the larger Hybrid-3 and Hybrid-4 ensembles, no metric showed a significant difference under the corrected test after Holm-Bonferroni correction (all Holm-adjusted p-values $\approx 1.0$; the smallest was $p = 0.4305$ for NLL in the Hybrid-2 vs. Hybrid-4 comparison). Adding a third or fourth model therefore produced no statistically detectable improvement over the two-model ensemble. Combined with the robustness and leave-one-metric-out analyses, this supports selecting the more parsimonious Hybrid-2 configuration, since the added complexity of larger ensembles is not justified by a measurable gain.

### 4.9 Final Ensemble Selection

Taken together, the results support Hybrid-2 as the final reliability-aware ensemble. Relative to the best individual model, Hybrid-2 improved the point estimates of the reliability metrics (reducing AURC by approximately 43%, NLL by approximately 17%, and worst-class ECE by approximately 36%) and improved the full eight-metric composite score. It was selected as the modal ensemble in 96.8% of random metric-weighting scenarios and remained unchanged across all eight LOMO ablations. Under the corrected repeated cross-validation paired t-test with Holm-Bonferroni correction, however, these per-metric improvements over the best individual model did not reach statistical significance (all Holm-adjusted $p \geq 0.168$); the case for Hybrid-2 therefore rests on its consistent directional gains together with its robustness and selection stability rather than on statistically significant per-metric differences.

Although Hybrid-3 and Hybrid-4 provided small additional improvements in selected metric point estimates, they did not differ significantly from Hybrid-2 on any metric after correction (all Holm-adjusted $p \approx 1.0$) and were less stable under both the Dirichlet robustness analysis and LOMO ablation. Therefore, Hybrid-2, composed of Swin-Tiny and TinyViT-5M, was selected as the final ensemble. This result demonstrates that reliability-aware ensemble selection can identify a parsimonious two-model configuration that improves reliability point estimates beyond the best individual classifier without relying on accuracy alone or increasing ensemble size unnecessarily.

### 4.10 Comparison with Representative SIPaKMeD Cervical Cytology Studies

Table 7 situates the proposed framework among representative deep learning studies that classify the SIPaKMeD cervical cytology dataset. The comparison is representative rather than a controlled head-to-head benchmark: the listed studies differ in their train–test splits, image preprocessing, data augmentation, training protocols, and the completeness with which they report performance and reliability metrics. Reported values are therefore taken from each source under its own experimental conditions. They are not directly commensurable, so the table is intended to characterize what each study evaluates rather than to rank methods by a single number. Only SIPaKMeD-based studies are included; papers evaluated solely on other cervical cytology datasets, such as Herlev or Mendeley LBC, are excluded so that the comparison remains confined to the SIPaKMeD benchmark on which the proposed framework is developed and evaluated.

***Table 7. Comparison of the proposed reliability-aware Hybrid-K framework with representative SIPaKMeD cervical cytology classification studies. Reported metrics are taken from each source under its own experimental protocol; "Not reported" indicates the metric was not provided in the original study. Bold row indicates the proposed study.***

| Study | Year | Dataset / task | Method / model | Validation protocol | Reported metrics | Reliability / statistical evaluation |
|---|---|---|---|---|---|---|
| Manna et al. [36] | 2021 | SIPaKMeD; five-class cervical cytology classification | Fuzzy rank-based CNN ensemble (Inception-v3, Xception, DenseNet-169) | Single split | Acc 0.954; F1 0.954; AUROC not reported | Calibration: not reported. Selective prediction/AURC: not reported. Statistical testing: no. Reliability-aware selection: no. |
| Demirtaş Alpsalaz et al. [65] | 2026 | SIPaKMeD; cervical cancer classification | Explainability-aligned reliability-weighted fuzzy ensemble | Single split | Acc 0.940; Macro-F1 0.940; AUROC 0.990 | Calibration: ECE, Brier, NLL (with post-hoc temperature scaling). Selective prediction/AURC: not reported. Statistical testing: no. Reliability-aware selection: partial (reliability-weighted fuzzy ensemble, not full multi-metric Hybrid-K selection). |
| Hu et al. [67] | 2026 | SIPaKMeD; cervical cytology classification | Quantile-deviation uncertainty ensemble with MLP meta-learner | Single split | Acc 0.981; F1 0.977; per-class AUROC 0.99–1.00 (ROC curves, first fold) | Calibration: not reported. Selective prediction/AURC: uncertainty considered, but AURC not reported. Statistical testing: yes (McNemar test). Reliability-aware selection: partial (uncertainty-aware ensemble, no AURC/composite robustness framework). |

| Study | Year | Dataset / task | Method / model | Validation protocol | Reported metrics | Reliability / statistical evaluation |
|---|---|---|---|---|---|---|
| Sangare et al. [68] | 2026 | SIPaKMeD; cervical cytological cell classification | ECAViT-Net: lightweight hybrid CNN–Transformer | 5-fold CV | Acc 0.9651; F1 0.9652; AUROC 0.9970 | Calibration: not reported. Selective prediction/AURC: not reported. Statistical testing: no. Reliability-aware selection: no. |
| Kılıç & Kılıç [69] | 2026 | SIPaKMeD; cervical cancer classification | DeepInsight-Net: CBAM-enhanced ResNet50 with focal loss | 5-fold CV | Acc 0.9963; F1 0.9874; per-class AUROC 0.99–1.00 (ROC curves) | Calibration: not reported. Selective prediction/AURC: not reported. Statistical testing: yes ($p < 0.001$). Reliability-aware selection: no. |
| **Proposed study** | **2026** | **SIPaKMeD; five-class cervical cytology classification** | **Reliability-aware Hybrid-K soft-voting ensemble (Swin-Tiny + TinyViT-5M)** | **Fixed stratified 5-fold partition × 3 training seeds; 15 paired fold-by-seed evaluations** | **Acc 0.9846; F1 0.9847; AUROC 0.9992** | **Calibration: ECE, WC-ECE, Brier, NLL. Selective prediction/AURC: yes (AURC / risk–coverage analysis). Statistical testing: yes (corrected repeated CV paired t-test with Holm-Bonferroni; no per-metric difference significant after correction). Reliability-aware selection: yes (eight-metric composite, Dirichlet robustness, and LOMO sensitivity).** |

***Note.*** Acc = accuracy; F1 = macro-averaged F1; AUROC = area under the receiver operating characteristic curve; ECE = expected calibration error; WC-ECE = worst-class ECE; NLL = negative log-likelihood; AURC = area under the risk–coverage curve; MLP = multilayer perceptron; CV = cross-validation. Values are as reported in each cited work and are not directly comparable across studies because splits, preprocessing, augmentation, and training protocols differ.

Overall, prior SIPaKMeD studies primarily emphasize discrimination performance, usually accuracy, F1, or AUROC. In contrast, the proposed Hybrid-K framework jointly evaluates discrimination, calibration, probabilistic reliability, selective prediction, statistical significance, and robustness to metric weighting. Although some individual studies report marginally higher headline accuracy or F1 (for example, [69] at 0.9963 accuracy), these are single-run point estimates obtained under heterogeneous protocols on a near-saturated benchmark and are not directly comparable to the multi-seed, statistically evaluated estimates reported here. The proposed framework attains comparable discrimination (accuracy 0.9846, macro-F1 0.9847, AUROC 0.9992) while additionally quantifying the calibration, selective-prediction, and weighting-robustness dimensions that these studies leave unaddressed; the contribution is therefore the breadth and robustness of reliability-aware selection rather than a marginal gain on any single discrimination metric.

## 5. Conclusion

This paper introduced a reliability-aware Hybrid-K ensemble selection framework for cervical cytology classification that explicitly incorporates calibration, probabilistic reliability, and selective prediction alongside discrimination when ranking candidate models and constructing ensembles. Nine candidate architectures spanning CNNs, vision transformers, and hybrid CNN-transformer designs were trained on the SIPaKMeD dataset under a shared protocol and ranked using an equal-weight, eight-metric composite score. The resulting Hybrid-2 ensemble, combining Swin-Tiny and TinyViT-5M via soft voting, reduced AURC by 43%, NLL by 17%, and worst-class expected calibration error by 36% relative to the best individual model, while matching or exceeding its classification performance.

Beyond raw performance gains, Hybrid-2's selection was shown to be robust to alternative metric weightings rather than an artifact of a single arbitrary weighting: it was selected as the modal ensemble in 96.8% of 5,000 randomly sampled Dirichlet metric-weight vectors, and its membership remained unchanged across all eight leave-one-metric-out ablations. Larger ensembles (Hybrid-3, Hybrid-4) provided only marginal additional gains on the composite score and were less stable under both robustness analyses, supporting Hybrid-2 as a parsimonious choice rather than a maximal one.

Several limitations should be considered when interpreting these results. The reliability-aware composite score used equal metric weights as its primary ranking criterion; while robustness to alternative weightings was explicitly evaluated, other weighting schemes reflecting different clinical priorities were not explored. It is also worth noting that the reliability metrics underlying the composite score are themselves imperfect proxies for true predictive trustworthiness; recent work has questioned how reliably common reliability metrics actually characterize model trustworthiness in medical imaging [73], suggesting that the composite ranking used here, while more comprehensive than single-metric selection, should not be treated as a definitive measure of clinical trustworthiness. A related limitation concerns the post-hoc calibration procedure itself. Temperature scaling was fitted on pooled held-out predictions from the remaining fold-level evaluations rather than on a calibration set withheld from training, and because the five-fold partition was reused across the three training seeds, the fitting data for each fold included the seed replicates of that same fold. The decision of whether to retain the fitted temperature was additionally made by comparing expected calibration error on the evaluation fold itself. The calibration data and the held-out evaluation fold are therefore not fully independent, and this dependence is expected to bias the calibration figures optimistically, most directly for ECE. Calibration-related quantities reported in this work, including ECE, WC-ECE, Brier score, and NLL, should accordingly be interpreted as internal exploratory estimates rather than as fully independent calibration-validation estimates, and the same qualification extends to the composite score and to the ensemble comparisons that depend on it. Re-estimating calibration on a strictly independent inner split, or restricting the fitting data to folds disjoint from the evaluation fold, is a necessary step before these calibration gains are treated as validated. A further limitation concerns the unit at which cross-validation was performed. Folds were constructed over individual cropped single-cell images without enforcing parent-image grouping, so the splitting procedure permits cells segmented from the same parent cluster image to fall into different folds. Because parent-image identifiers are recoverable from the SIPaKMeD filenames, this reflects a design choice in the present partitioning rather than a constraint of the data. The reported estimates should therefore be read as image-level internal validation, and they may be optimistic relative to what would be obtained under parent-image-, slide-, or specimen-level partitioning. Group-aware cross-validation using the parent-image identifiers, which would quantify the magnitude of this effect directly, is left to future work. Most importantly, although SIPaKMeD provides a well-established benchmark for five-class cervical cytology classification, the present study is limited to internal validation on this dataset. External validation on larger, independently collected, taxonomy-compatible single-cell cervical cytology datasets remains an important direction for future work, particularly for assessing calibration and selective-prediction behavior under acquisition and population shifts. This direction is motivated by a broad, well-documented body of evidence that distribution shift degrades deep learning performance in medical imaging: cross-population and cross-scanner shifts have been shown to reduce performance across imaging modalities [74], physical imaging parameter variation alone can drive substantial domain shift [75], and scanner-induced shift in particular has been shown to affect deep learning performance even when models are trained with explicit transfer learning [76]. Evidence that calibration and related reliability properties can transfer decoupled from discrimination across cohorts in clinical prediction [77] further underscores that reliability gains established through internal cross-validation should not be assumed to carry over to a new domain without direct external assessment.

Future work should pursue external validation on larger, independently collected, taxonomy-compatible single-cell cervical cytology datasets and additional imaging protocols, re-partition the dataset using group-aware cross-validation keyed on the parent-image identifiers encoded in the SIPaKMeD filenames, so that all cells segmented from a given parent image are confined to a single fold, explore learned or clinically elicited metric weightings in place of the equal-weight default, and investigate how selective prediction and referral thresholds interact with ensemble composition. More broadly, the reliability-aware selection framework introduced here could be applied beyond cervical cytology to other medical image classification tasks where calibrated, selectively deployable models are needed. Beyond model development, realizing these reliability gains in practice will require attention to how such models are integrated into clinical workflows and health information systems; human-centered studies of technology design and adoption in complex-care settings provide useful guidance for this translational step [78-80]. Taken together, this work demonstrates that reliability-aware ensemble selection can identify a compact ensemble that is robust to alternative metric weightings and improves calibration and selective-prediction point estimates without sacrificing classification accuracy, providing a rigorous internal foundation on SIPaKMeD that motivates subsequent external validation before such reliability gains are assumed to generalize to new clinical settings.

## Declarations

**Funding:** This research received no external funding.

**Institutional Review Board Statement:** Not applicable. This study used a publicly available, de-identified dataset and did not involve direct human participant recruitment or intervention.

**Informed Consent Statement:** Not applicable. This study used a publicly available, de-identified dataset and did not involve direct interaction with human participants.

**Data Availability Statement:** The SIPaKMeD cervical cytology dataset used in this study is publicly available from its original source.

**Conflicts of Interest:** The authors declare no conflicts of interest.